\documentclass[12pt]{article}
\usepackage[export]{adjustbox}
\usepackage{authblk}
\usepackage{bbm}
\usepackage [autostyle, english = american]{csquotes}
\MakeOuterQuote{"}
\usepackage{pgfpages}
\usepackage{graphicx}
\usepackage{multicol}
\usepackage{csquotes}
\usepackage{multirow}
\usepackage{graphicx} %\% Allows including images
\usepackage{booktabs} % Allows the use of \toprule, \midrule and \bottomrule in tables
\usepackage{authblk}
\usepackage{caption}
\usepackage{subcaption}
\usepackage{stackengine}
\usepackage{amsmath}
\usepackage{comment}
\usepackage{hyperref}
\hypersetup{
    colorlinks=true,
    linkcolor=blue,
    filecolor=blue,      
    urlcolor=blue,
    citecolor=blue,
}
\usepackage[spanish,english]{babel}

\usepackage{amssymb}
\usepackage{stackrel}
\usepackage{natbib} % bibliography

\usepackage[capposition=top]{floatrow}

\usepackage{lscape}

\usepackage[margin=1in]{geometry}

\title{Granular Linkages, Supplier Cost Shocks \\ \& \\ Export Performance}

\author[1]{Santiago Camara}
\affil[1]{Northwestern University \& Red-NIE}
\date{\today}

\begin{document}

\maketitle

\begin{abstract}
    This paper presents evidence on the granular nature of firms' network of foreign suppliers and studies its implications for the impact of supplier shocks on domestic firms' performance. To demonstrate this, I use customs level information on transactions between Argentinean firms and foreign firms. I highlight two novel stylized facts: (i) the distribution of domestic firms' number of foreign suppliers is highly skewed with the median firm reporting linkages with only two, (ii) firms focus imported value on one top-supplier, even when controlling for firm size. Motivated by these facts I construct a theoretical framework of heterogeneous firms subject to search frictions in the market for foreign suppliers. Through a calibration exercise I study the framework’s predictions and test them in the data using a shift-share identification strategy. Results present evidence of significant frictions in the market for foreign suppliers and strong import-export complementarities.
    
    \medskip
    \medskip

\end{abstract}

\newpage
\section{Introduction} \label{sec:introduction}

Emerging Market economies rely heavily in imported intermediate inputs and capital goods (see \cite{eaton2001trade}, for example). The aggregate import flow of intermediate inputs and capital goods is composed of millions of transactions between domestic individual firms and foreign suppliers. A large share of these import-transactions are carried out by exporting firms which are highly dependent on foreign imported inputs (see \cite{brambilla2012exports,kasahara2013productivity}). While the rise of a literature interested in firm heterogeneity has pointed out the importance of differences in exporters (see \cite{bastos2018export}) and importers  (see \cite{blaum2018gains}) there is still little evidence on firms' differences in their network of foreign suppliers. Additionally, there is little evidence on how shocks or disruptions in international supplier-linkages impact domestic firms' performance. This is mostly due to the fact that customs level data sets are usually aggregated to the level of the individual domestic firm, summed across all foreign exporters. In this paper I present novel stylized facts about Argentinean firms' network of foreign suppliers, emphasizing its granular nature and the presence of search and informational frictions. Furthermore, I show that a shock which increases the price a foreign supplier charges for imported inputs leads to a drop in domestic firms' export performance.

\begin{figure}[ht]
    \centering
    \caption{Relevance of Imported Inputs for Production }
    \label{fig:impo_introduction}
    \begin{subfigure}[b]{0.45\textwidth}
    \centering    
    \includegraphics[scale=0.5]{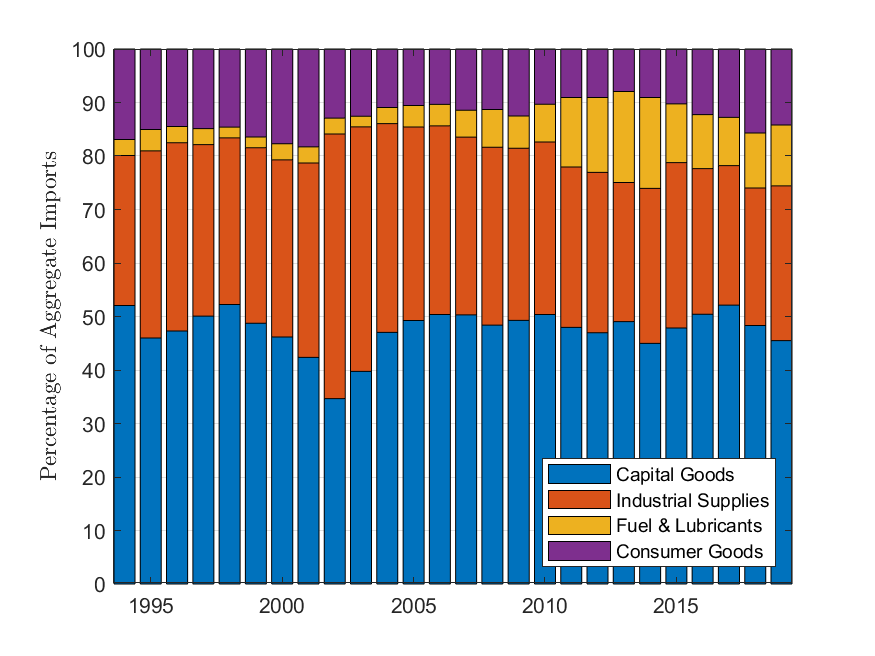}
    \caption{Import decomposition - Main end use}
    \label{fig:impo_bec}
     \end{subfigure}
     \hfill
    \begin{subfigure}[b]{0.45\textwidth}
    \centering
    \includegraphics[scale=0.5]{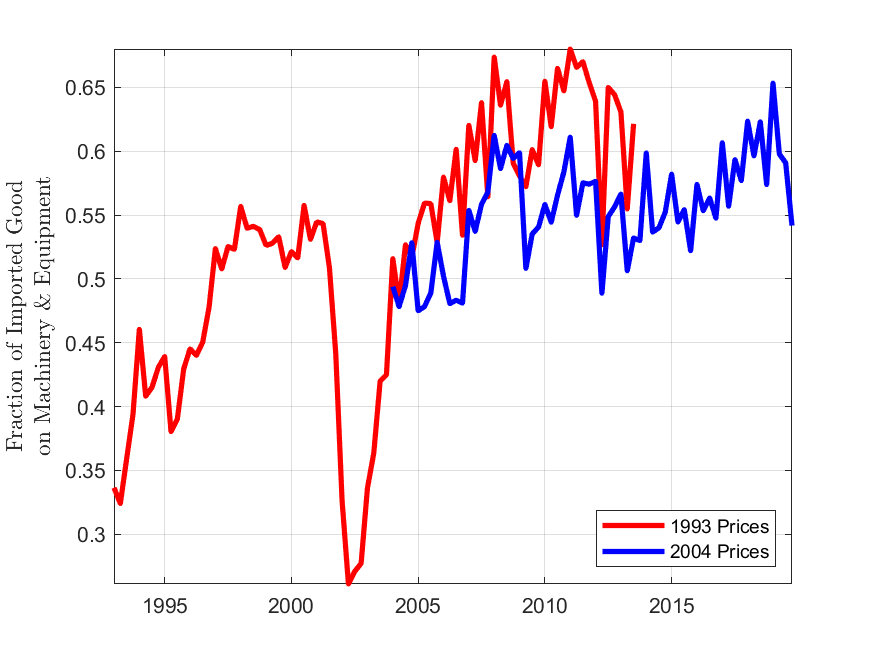}
    \caption{Share of Imports in Investment}
    \label{fig:impo_investment}
     \end{subfigure}
     \floatfoot{\footnotesize \textbf{Note:} The left panel presents the decomposition of total imported value according to the main end use of goods, using the latest revision of the Broad Economic Category (BEC) classification. The right panel presents data on the fraction of imported goods on total investment in capital, sourced from Argentina's national accounts.}
\end{figure}

I carry out this analysis by using a unique data set of Argentinean firms that identifies both the domestic importing firm and the foreign exporting supplier. The reliance of Argentinean firms on foreign intermediate and capital goods can be observed at the aggregate level in Figure \ref{fig:impo_introduction}. On the left panel, Figure \ref{fig:impo_bec} shows that the import of intermediate and capital goods explain between 80\% and 90\% of Argentina's total imports for the period 1994-2019. On the right panel, Figure \ref{fig:impo_investment} shows that imported capital goods explain between 55\% and 65\% of the total investment in machinery and/or equipment for the same time period.
These aggregate facts motivate the analysis of firms network of foreign suppliers and to study how supplier-specific shocks may affect domestic firms' performance.

I begin my analysis by documenting two stylized facts about domestic firms' network of foreign suppliers. First, I show that the distribution of number of foreign suppliers per domestic firm is highly skewed towards high values, with the median and mean number of foreign suppliers equal to 2 and 6.11 respectively. This fact suggest that for a vast majority of Argentinean firms their network of foreign suppliers is granular and not  diversified. Additionally, I show that the right tail of the distribution of number of suppliers is primarily explained by exporting firms which rely heavily on foreign inputs. Second, I show that even when controlling for firms' size and/or export status, firms tend to focus most of their imported value on one main foreign supplier. This results provides additional evidence that domestic firms' network of foreign suppliers is granular and that firms are exposed to shocks or disruptions in their linkages with foreign suppliers. 

Then, I complement the stylized facts on firms' granular networks of foreign suppliers with additional information which suggest the presence of search and informational frictions in the market for these foreign suppliers. I show that a vast majority of importing firms establish new linkages with foreign suppliers every year, albeit only explaining a small share of aggregate imports. Furthermore, I show that firms' supplier sets experience high turn-over with only their top-supplier relationship, i.e. the supplier which explains the highest share of imported value, exhibiting persistence in time. I argue that these facts provide evidence of the presence of search and informational frictions in international markets.\footnote{This is in line with a previous literature in international trade and in labor economics.}

Next, motivated by these facts, I build a heterogeneous firm model with trade in intermediate inputs, subject to search and matching frictions in the market for foreign suppliers and selection into exporting. Foreign suppliers are heterogeneous in their efficiency to provide a differentiated intermediate input at a low price. While firms have incentives to match with the most efficient set of foreign suppliers, searching involves the payment of a fixed cost and matching is random, i.e., firms cannot choose with which foreign supplier to match. Through calibration and numerical exercises, I show how the model yields two sharp predictions with respect to the impact of a negative supplier-supply shock which rises domestic firms' cost of imported inputs: (i) the presence of strong import-export complementarities, (ii) relatively smaller firms are hit hardest as they have smaller, less diversified network of foreign suppliers. Furthermore, I argue that the impact of this supplier-supply shock is increasing in the severity search and information frictions in the market for foreign suppliers. 

Finally, I take my theoretical framework to the data and test its predictions. To do so, I construct firm level import supplier-supply shocks using a shift-share analysis. I argue that this empirical strategy is consistent with the search and matching frictions introduced in the theoretical framework. This identification strategy exploits firms' lagged exposure to foreign suppliers through imported-input linkages and supplier-time specific changes in import prices as shifts. I find that a one standard deviation firm level shock decreases imported and exported quantities by 0.79\% and 0.58\% respectively, suggesting strong complementarities between imported inputs and exported performance. Furthermore, the estimated impact is larger for relatively smaller firms is between two and three times bigger than the benchmark full sample estimate.

\noindent
\textbf{Related literature.} This paper relates primarily to three strands of literature. First, it relates to an incipient literature which studies linkages or connections between firms in different countries. For instance, \cite{bernard2018networks} characterize the connections between Norwegian and foreign firms and show that most domestic firms have few connections with foreign firms while a small number of firms have many connections. This is a stylized fact also highlighted by \cite{atalay2011network} for firm-to-firm connections within the US market, by \cite{tintelnot2018trade} for the Belgium economy, and by \cite{bernard2018firm} for Colombian firms' linkages with foreign suppliers. This paper contributes to this literature by describing both the domestic firms' number of linkages with foreign suppliers and the sourcing strategy within those suppliers as well. In particular, I find that even controlling for firms' export status and employment, most firms focus a large fraction of their imported value on one single foreign supplier. This is in line with French firms' highly concentrated portfolio of export partners as shown by \cite{kramarz2020volatility}.

Second, this paper relates to literature in international trade which considers how foreign trade shocks matter for import and export performance in the presence of search and matching frictions. For instance, \cite{bernard2018two} shows that worsened market access during the Great Financial Crisis of 2008-2009 had a significant negative impact on production costs, especially for downstream firms that were \textit{ex-ante} highly exposed to international markets. Another example is \cite{kramarz2020volatility} which provides evidence that even the largest French firms' have a concentrated portfolio of export customers. Additionally, the authors show through a model of trade networks, that concentration of export flows implies that individual exporters are strongly exposed to microeconomic demand shock which does not average out in the aggregate. This paper's contribution to this literature is two-fold: (i) by providing evidence of domestic firms' high concentration of imported value on one main foreign suppliers, (ii) by showing through a shift-share analysis that supplier cost shocks have a significant impact in domestic firms' import and export performance. I argue that these results are evidence that the granular nature of firms' network of foreign suppliers leaves them exposed to idiosyncratic supplier-supply shocks.

Third, this paper relates to literature in international trade which examines the complementarities between imported inputs and export performance. \cite{kasahara2013productivity} showed that larger and more productive firms tend to both import and export more. Furthermore, the authors show that due to import and export complementarities, increases in the cost of imported intermediate inputs can have a large adverse effect on the exports of final goods. Other related literature has also emphasized the importance of imported inputs for the production of high-quality goods. For instance, \cite{bastos2018export} show that firms which source higher-quality imported inputs tend to export higher quality products to richer and more selective countries.\footnote{For other examples which present evidence of complementarities between imported inputs and export performance see \cite{fan2015trade} and \cite{bas2015input}.} Closely related to this paper, \cite{camara2022does} shows that exports of differentiated goods, particularly to high-income countries, are more sensitive to increases in the cost of imported inputs than exports of commodity-based goods. This paper contributes to this literature by studying how the granular nature of firms' network of foreign suppliers matters for the transmission of shocks and their impact on export performance. In addition, this paper argues that relatively smaller firms, which can only develop small-scale networks of foreign suppliers, are hit particularly hard.

\noindent
\textbf{Organization.} This paper is comprised of \ref{sec:conclusion} sections, starting with the present introduction. Section \ref{sec:data_facts} describes the data sets used in this paper and presents novel stylized facts on domestic firms' network of foreign suppliers. Section \ref{sec:model} presents a heterogeneous firm model subject to search and matching frictions and describes its predictions. Section \ref{sec:supplier_shocks} presents empirical results on the impact of a supplier import price shock on export performance. Section \ref{sec:conclusion} concludes.

%%%%%%%%%%%%%%%%%%%%%%%%%%%%%%%%%%%%%%%%%%%%%%%%%%%%%%%%%%%%%%%%%
\section{Data Description \& Stylized Facts} \label{sec:data_facts}

In this section I describe the main datasets used in this paper and present novel stylized facts about domestic firms' network of foreign suppliers. Section \ref{subsec:data_description} describes the firm-to-firm dataset and complementary datasets used, and briefly describes any dataset cleaning procedures. Section \ref{subsec:stylized_facts} presents the two novel stylized facts describing the granular nature of firms' network of foreign suppliers. Section \ref{subsec:search_evidence} complements these novel facts by providing evidence of search and matching frictions in the market for foreign suppliers.

\subsection{Data Description} \label{subsec:data_description}

First, I use data from administrative customs records for Argentinean firms. It is based on customs declarations forms collected by Aduana Argentina (Argentina's customs office), and comprises the entire universe of Argentinian trade flows for the period 2000-2008. Information is dis-aggregated by firm, date of shipment frequency, product at 6-digit HS good classification, exports by destination, and imports by source country.\footnote{Argentinean customs provide a product description which classifies goods at 12 digit level. This classification builds on on the 6-digit HS international good classification. The first two extra digits are common to a MERCOSUR nomenclature. The last four digits (actually three digits and a letter) is Argentina specific. However, there is no concordance table across time for this classification. Consequently, using this classification could lead to spurious results because of attrition.}$^{,}$\footnote{Given that these datasets are constructed using declaration forms prone to mistakes, I drop observations which do not report and importer or exporter identifier, a valid HS6 or country code for the product transacted, and for trade flows under USD\$1,000.} In Appendix \ref{sec:appendix_features_data}, I present summary statistics on the distribution of firms' export and import per firm, which provides additional details on these datasets used across the paper.

For import flows, the data set allows me to observe the foreign supplier-exporting firm. The foreign supplier ID recovered from the customs declarations is the self-declared name of the company, i.e., not a state-issued tax ID number. Since foreign suppliers do not have a unique ID that identifies them, their registration is not harmonized and is subject to errors. Therefore, I carry out a cleaning procedure that allows me to join the information of suppliers that were registered in two or more different ways. Details of this cleaning procedure follow the methodology used by \cite{berninietal2021} and are described in Appendix \ref{subsec:Appendix_cleaning}. 

Second, I have additional firm level information to complement the international trade dataset. In particular, I count with firm level data on firms' number of employees. The source of this data is SIPA or ``Sistema Integrado Previsional Argentino'' and the time lapse is 2001-2008. The coverage of the dataset on firm level employment differs across time but firms in this data set represents on average above 98\% of aggregate import and export value. Results presented in Section \ref{sec:supplier_shocks} hold for both the full sample of importers and exporters and for the sub-sample for which I have additional employment data. Additional details on the coverage of these dataset can be found in Appendix \ref{sec:appendix_features_data}.

%%%%%%%%%%%%%%%%%%%%
\subsection{Firm-to-Firm Stylized Facts} \label{subsec:stylized_facts}

Next, I present key stylized facts which describe Argentinean firms' network of foreign suppliers. First, I show domestic firms' network of foreign suppliers is granular, with the median domestic firm purchasing from only two foreign suppliers. Second, I argue that even large firms which purchase intermediate inputs from a relatively higher number of foreign suppliers, still source most of their imports from one main supplier. I argue that these facts provide evidence that domestic firms are highly exposed to supplier-specific cost shocks.

\begin{figure}[ht]
    \centering
    \caption{Domestic Firms' Number of Foreign Supplier \\ \small  Histograms}
    \label{fig:Histogram_Number_Suppliers}
     \centering
     \begin{subfigure}[b]{0.495\textwidth}
         \centering
         \includegraphics[width=\textwidth]{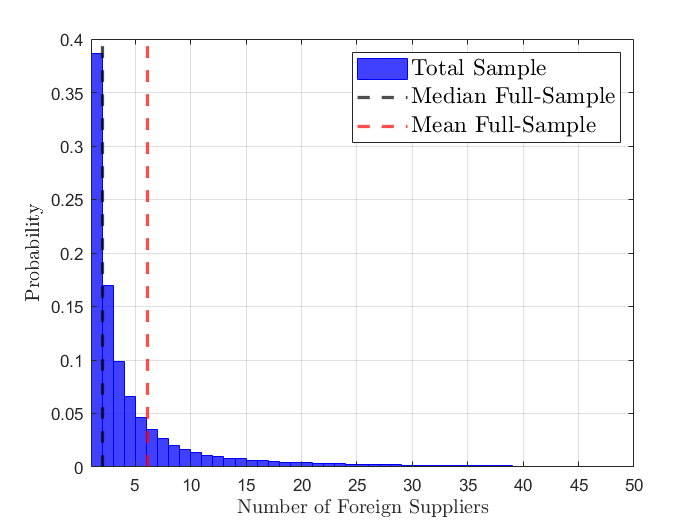}
         \caption{Full Sample}
         \label{fig:histogram_number_suppliers_full_sample}
     \end{subfigure}
     \hfill
     \begin{subfigure}[b]{0.495\textwidth}
         \centering
         \includegraphics[width=\textwidth]{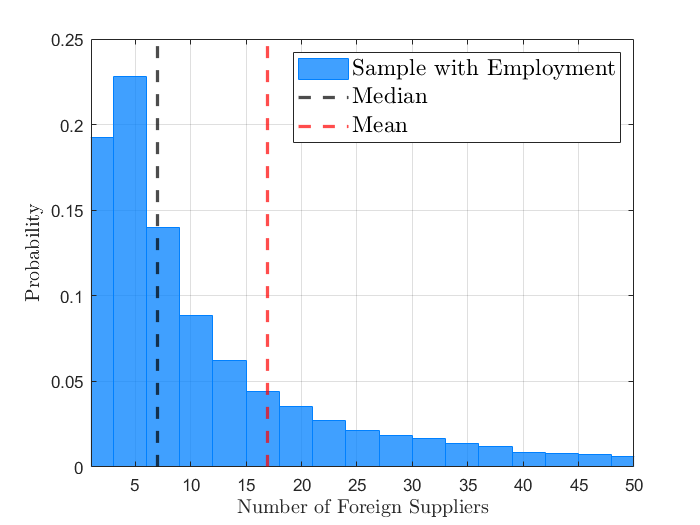}
         \caption{Sample with Employment}
         \label{fig:histogram_number_suppliers_full_employment}
     \end{subfigure} \\
     \begin{subfigure}[b]{0.495\textwidth}
         \centering
         \includegraphics[width=\textwidth]{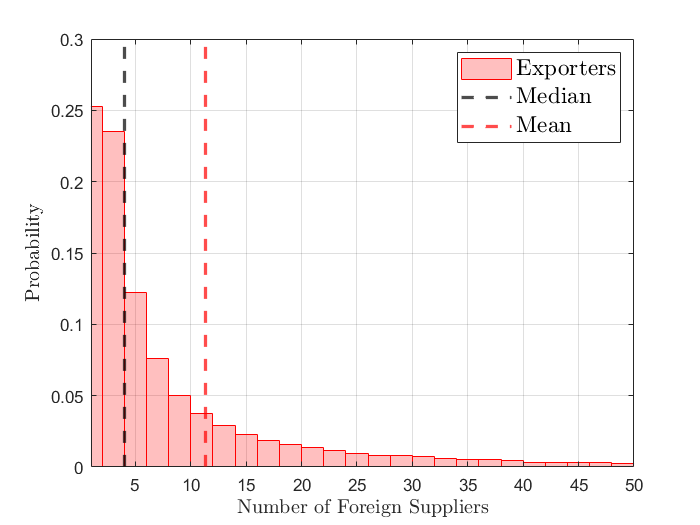}
         \caption{Exporting Firms}
         \label{fig:histogram_number_suppliers_exporters}
     \end{subfigure}
     \hfill
     \begin{subfigure}[b]{0.495\textwidth}
         \centering
         \includegraphics[width=\textwidth]{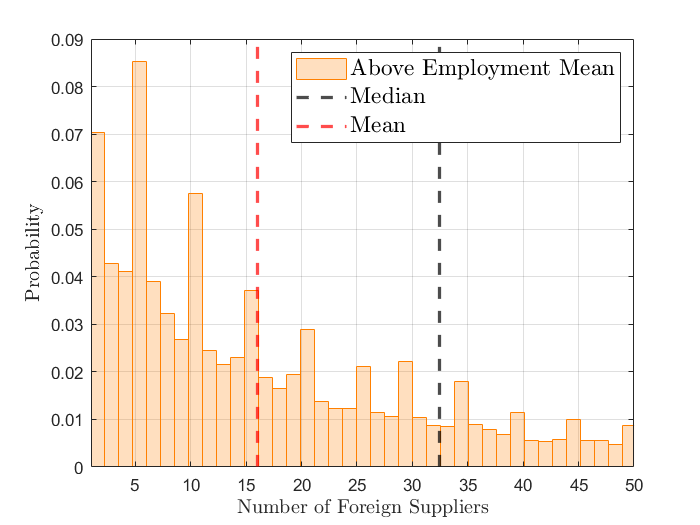}
         \caption{Above Mean Employment}
         \label{fig:histogram_number_suppliers_employment_above}
     \end{subfigure}
\end{figure}

I start by describing the distribution of domestic firms' number of foreign suppliers. Figure \ref{fig:Histogram_Number_Suppliers} presents the histogram of domestic firm's number of foreign suppliers for different sample partitions. 
First, on top left panel, Figure \ref{fig:histogram_number_suppliers_full_sample} shows the histogram for the full sample (in blue) and for exporting firms (in red). The median and mean number of foreign suppliers for the full sample is 2 and 6.11, respectively. This implies that the distribution is highly skewed towards higher values (Table \ref{tab:appendix_dist_number_suppliers} in Appendix \ref{sec:appendix_stylized_facts} presents more statistics on the distribution of domestic firms' number of foreign suppliers). These figures suggest that a vast majority of Argentinean firms source imports from a limited number of foreign suppliers, i.e., that firms' network of foreign suppliers is granular. On the top right panel, Figure \ref{fig:histogram_number_suppliers_full_employment} shows that for the subset of firms for which I have employment data, firms still exhibit granular networks of foreign suppliers, with the median and mean number of foreign suppliers equal to 7 and 16.90. \textit{A priori}, this granularity implies that a significant majority of firms are exposed to idiosyncratic supply shocks from their foreign suppliers.

Second, I turn to analyzing which type of firms expose the right tail of the distribution of number of suppliers. Figures \ref{fig:histogram_number_suppliers_exporters} and \ref{fig:histogram_number_suppliers_employment_above} show that exporting firms and relatively larger firms in terms of employment have a higher number of foreign suppliers, compared to the full sample. On the one hand, exporting firms have a median and mean number of suppliers of 4 and 11.33 respectively. On the other hand, firms which employment levels are above their sector's mean have a median and mean number of suppliers of 16 and 32.46 respectively. Furthermore, Table \ref{tab:appendix_dist_number_suppliers} in Appendix \ref{sec:appendix_stylized_facts} shows that firms which are above the mean level of exports per firm have a median and mean number of suppliers equal to 16 and 36.97, respectively. Overall, these results show that larger firms, which export higher amounts also import from a larger number of foreign suppliers. This result suggests the presence of strong complementarities between imported inputs and export performance.\footnote{This result is also in line with a \cite{kasahara2013productivity} who argues that firms which exhibit higher exported value also have a higher exported imported value.} 

\begin{figure}[ht]
    \centering
    \caption{Share of Imported Value explained by Top-Supplier \\ \small  Histograms}
    \label{fig:Histogram_Share}
     \centering
     \begin{subfigure}[b]{0.495\textwidth}
         \centering
         \includegraphics[width=\textwidth]{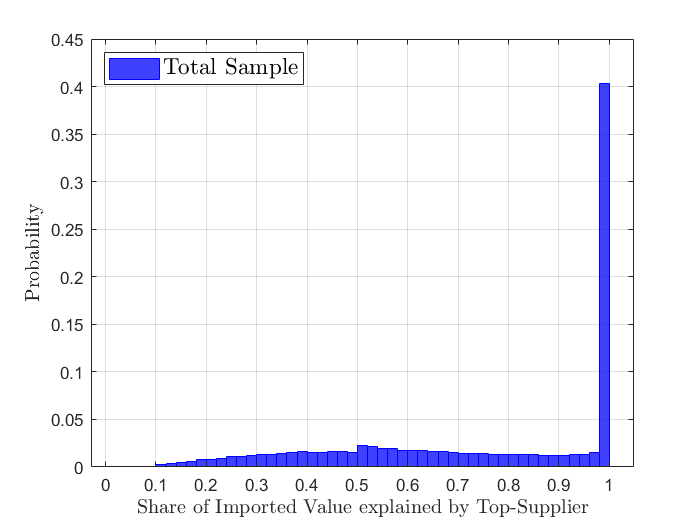}
         \caption{Full Sample}
         \label{fig:histogram_top_supplier_full_sample}
     \end{subfigure}
     \hfill
     \begin{subfigure}[b]{0.495\textwidth}
         \centering
         \includegraphics[width=\textwidth]{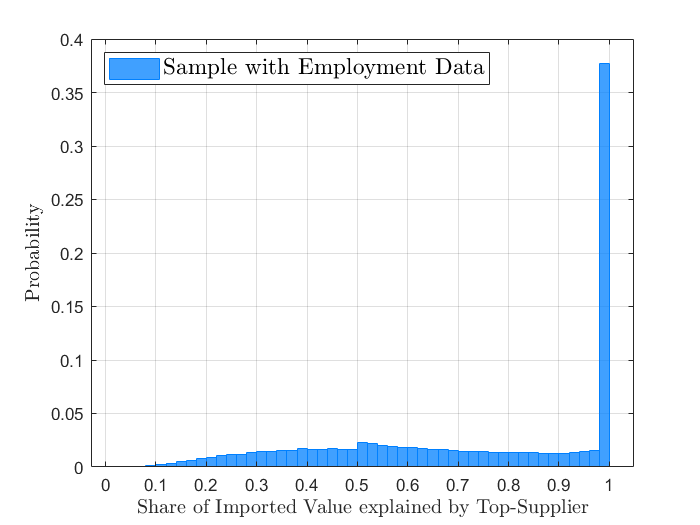}
         \caption{Sample with Employment Data}
         \label{fig:histogram_top_supplier_full_employment}
     \end{subfigure} \\
    \begin{subfigure}[b]{0.495\textwidth}
         \centering
         \includegraphics[width=\textwidth]{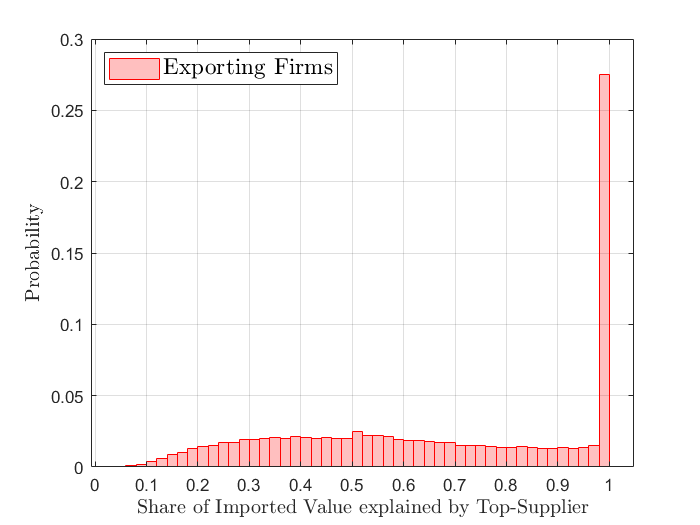}
         \caption{Exporting Firms}
         \label{fig:histogram_top_supplier_exporters}
     \end{subfigure}
     \hfill
     \begin{subfigure}[b]{0.495\textwidth}
         \centering
         \includegraphics[width=\textwidth]{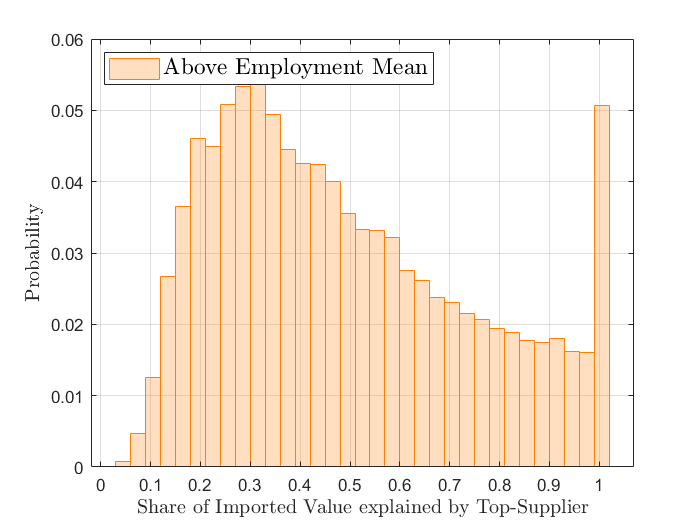}
         \caption{Above Mean Employment}
         \label{fig:histogram_top_supplier_employment_above}
     \end{subfigure}
\end{figure}

Next, I turn to analyzing the  distribution of domestic firms' imported value across their foreign suppliers. While Figure \ref{fig:Histogram_Number_Suppliers} shows that relatively larger firms have more linkages with foreign suppliers than relatively smaller firms, it says nothing about how firms allocate imported value across their foreign suppliers. One way to analyze domestic firms' degree of imported value concentration is to study the share explained by their most important or \textit{``top-supplier''}. Figure \ref{fig:Histogram_Share} presents histograms on the distribution of firms' share of imported value explained by their top-supplier.
On the top left panel, Figure \ref{fig:histogram_top_supplier_full_sample} shows that a large fraction of firms source the vast majority of their imported value from their top-supplier. In particular, the median and mean share of firm-level imported value explained by ``top-suppliers'' is 83.2\% and 74.6\%. This concentration of imported value on one supplier is an additional indicator that a large share of domestic firms have not well diversified network of foreign suppliers.

\textit{A priori}, one could think that this result is driven by relatively smaller firms which have a relatively low number of foreign suppliers. However, this is not the case. Figures \ref{fig:histogram_top_supplier_full_employment} and \ref{fig:histogram_top_supplier_exporters} show that for firms which have a larger number of foreign suppliers, such as firms for which I have employment data and for exporting firms, ``top-suppliers'' still explain the vast majority of firms' imported value.  While the histogram is less skewed for firms with levels of employment above their sector's mean, in Figure \ref{fig:histogram_number_suppliers_employment_above}, these firms still allocate a significant share of their imported value on their top-supplier. For this set of firms, 5\% source all of their imports from their top-supplier, the median and mean firm source 44.3\% and 49.3\% from their top-supplier and the tenth percentile sourcing 19.2\% from their top-supplier.\footnote{Table \ref{tab:appendix_dist_share_top} in Appendix \ref{sec:appendix_stylized_facts} presents additional information on the distribution of the share of imported value explained by firms' top-supplier across different sample partitions.} Thus, even relatively larger firms focus a significant share of their total imported value on their top-supplier. 

In summary, this section presented evidence on the granular nature of Argentinean firms' import sourcing strategies. First, I showed that the distribution of domestic firms' number of suppliers has low median and mean values (2 and 6.11, respectively) but significantly skewed towards higher values, primarily explained by large exporting firms. I showed that while these large and exporting firms source imports from a relatively higher number of foreign suppliers they still source most of their imported value from their top-supplier. I take these results as evidence that domestic firms' sourcing strategies are granular which exposes them to idiosyncratic supplier price shocks. 

\subsection{Evidence of Search \& Matching Frictions} \label{subsec:search_evidence}

I complement the facts presented above by providing evidence that domestic firms face search and matching frictions in the market for foreign suppliers. To this end, I characterize the frequency with which firms search for new suppliers and the survival profiles of the linkages between domestic and foreign suppliers.

\begin{figure}[ht]
    \centering
    \caption{Evidence on Firms' Searching for New Suppliers}
    \label{fig:Search_Evidence}
     \centering
     \begin{subfigure}[b]{0.495\textwidth}
         \centering
         \includegraphics[width=\textwidth]{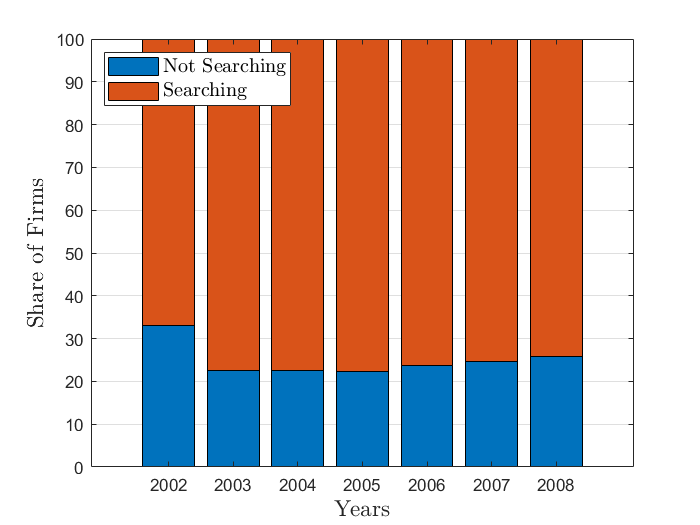}
         \caption{Share of Firms Searching}
         \label{fig:share_firms_searching}
     \end{subfigure}
     \hfill
     \begin{subfigure}[b]{0.495\textwidth}
         \centering
         \includegraphics[width=\textwidth]{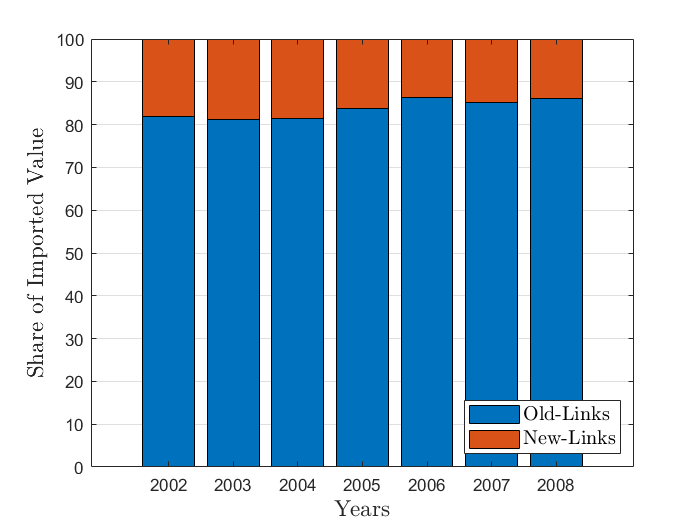}
         \caption{Share of Total Imports explained by Search}
         \label{fig:share_searching}
     \end{subfigure}
\end{figure}

To begin with, I show that a significant share of importing firms purchase goods from a new foreign supplier every year. I label firm $i$ as ``searching'' in year $t$ if I observe an import transaction between her and a foreign supplier $j$ which I have not observed in firm $i$'s supplier set in previous years. Given that the sample covers the years 2000-2008, there is a risk of misidentifying linkages established in periods before the year 2000 as new linkages. To minimize this risk, I identify searching firms starting from year 2002 onward. Figure \ref{fig:Search_Evidence} presents evidence on the importance of searching for new foreign suppliers in terms of the share of importing firms and in terms of total aggregate imports. On the left panel, Figure \ref{fig:share_firms_searching} shows that out of the total number of importing firms, on average, close to 75\% of them exhibit a new linkage with a foreign supplier every year.\footnote{Note that searching firms in period $t$ include both firms which had already imported in periods before $t$ and firms which start importing for the first time in period $t$. A firm which imports for the first time in period $t$ is by default searching as it has never established a linkage with a foreign supplier before.} The share of searching firms is relatively stable across the period 2000-2008.
On the right panel, \ref{fig:share_searching} presents the share of total value imported at the aggregate level explained by domestic firms searching for new suppliers. This figure suggests that while a significant share of firms search every period for new suppliers, newly established linkages or ``\textit{new-links}'' only explain between 15\% and 20\% of total value imported. This is, the vast majority of total value imported is explained by already established or ``\textit{old-links}'' between domestic firms and foreign suppliers.

I turn to analyzing the survival profile of linkages between domestic firms and foreign suppliers. In particular, I compute the probability that a linkage between a domestic firm $i$ and a foreign supplier $j$ active in period $t$ is also active in period $t+s$, for $s \geq 1$.
\begin{figure}[ht]
    \centering
    %\caption{Domestic Firm - Foreign Supplier Survival Profiles \& Persistence}
    \caption{Survival Profile by Type of Linkage}
    \label{fig:survival_noFE}
         \centering
         \includegraphics[width=12cm,height=9cm]{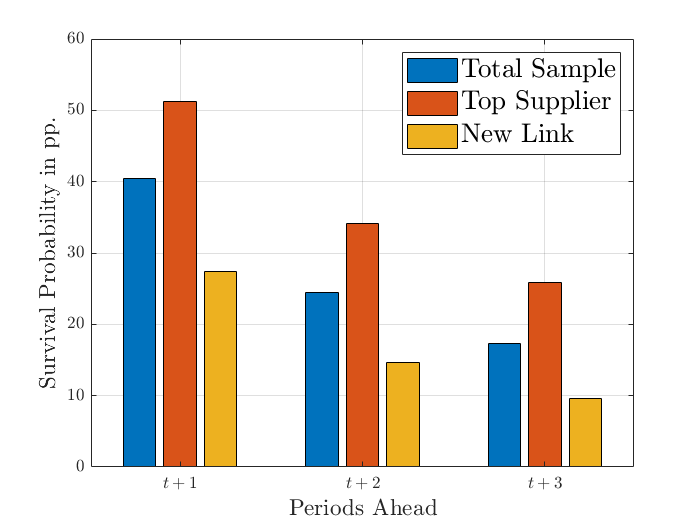}
    \floatfoot{\footnotesize \textbf{Note:} The survival probability is computed through a regression exercise. For the total sample result, the regression $\mathcal{P}_{i,j,t+s} = \beta_0 + \epsilon_{i,j,t+s}$ is estimated with, where $\mathcal{P}_{i,j,t+s}$ is an indicator function which takes the value of 0 if the link is not active in period $t+s$ and takes the value of 1 if the link is active. Parameter $\beta_0$ is reported. For the case of ``Top-Supplier'' and ``New Link'' the regression $\mathcal{P}_{i,j,t+1} = \beta_0 + \beta_1 \text{Type-Link}_{i,j,t} + \epsilon_{i,j,t+s}$, where $\text{Type-Link}_{i,j,t}$ is an indicator function of the type of linkage, top supplier or new linkage, and the sum of coefficients $\beta_0$ and $\beta_1$ is presented. For additional details and results on these regressions see Tables \ref{tab:SF_survival} and \ref{tab:SF_survival1_entrant} in Appendix \ref{sec:appendix_stylized_facts}.}
\end{figure}
Figure \ref{fig:survival_noFE} presents the survival probability in periods $t+s$ for $s=\{1,2,3\}$ for three types of linkages between domestic firms and foreign suppliers (i) total sample or all linkages (blue bar), (ii) top supplier or the linkage between a domestic firm and her supplier which explains the highest share of total value imported (red bar), (iii) new link or a newly established linkage (yellow bar). The total sample survival probability is close to 40\% one period ahead, 25\% two periods ahead, and 18\% three periods ahead. These survival probabilities are significantly low, compared to firms' survival probabilities in domestic markets, and in line with low survival probabilities in export markets, see \cite{albornoz2012sequential,fanelli2015export,eaton2021search}.\footnote{In Appendix \ref{sec:appendix_features_data} I show that the survival probability of a firm in Argentina of remaining active one and two years ahead in the domestic market is 96\% and 92\% respectively.} These papers have also suggested that low survival probabilities have been associated as evidence of search and information frictions in international markets.\footnote{Similarly, the literature in labor markets have used high job and worker turnover in labor markets as evidence of search and informational frictions (see \cite{cooper2007search,pissarides2011equilibrium}).}  Thus, the low survival profile combined with firms' high rate of search in international markets suggest that firms' supplier set experiences high turnover.

Lastly, Figure \ref{fig:survival_noFE} also presents evidence of significant heterogeneity in survival profiles across types of linkages. On the one hand, the survival profile of firms' new links (yellow bar) is between 40\% and 50\% lower than the survival profile of the total sample (red bar). On the other hand, the survival profile of firms' top-supplier (red bar) is between 25\% and 35\% higher than the total sample survival profile (blue bar). This result is consistent with the theory that domestic firms face informational frictions in the market for foreign suppliers. \textit{A priori}, one would expect that a domestic firm has better knowledge on the efficiency of her top-supplier than that of a newly linked supplier. Thus, it would be more likely for a domestic firm to break a linkage with a newly linked supplier than to break a linkage with its top supplier.

Lastly, I provide supporting evidence on the persistence of firms' top-supplier linkage.
\begin{figure}[ht]
    \centering
    \caption{Persistence of Top Supplier Linkages \\ \footnotesize Probability of remaining as Top Supplier}
    \label{fig:persistence_top_supplier}
     \centering
     \begin{subfigure}[b]{0.495\textwidth}
         \centering
         \includegraphics[width=\textwidth]{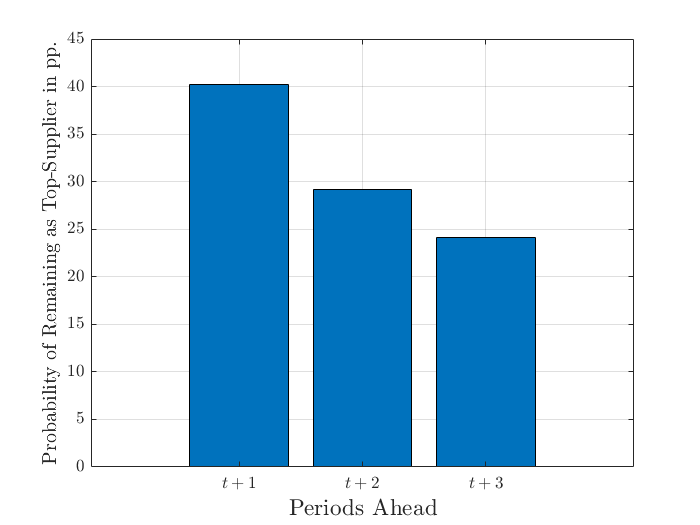}
         \caption{Unconditional Probability}
         \label{fig:persistence_unconditional}
     \end{subfigure}
     \hfill
     \begin{subfigure}[b]{0.495\textwidth}
         \centering
         \includegraphics[width=\textwidth]{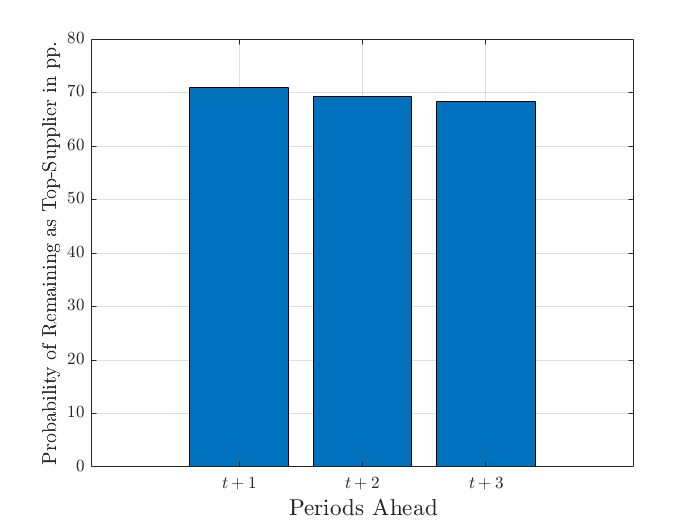}
         \caption{Conditional Probability}
         \label{fig:persistence_conditional}
     \end{subfigure}
    \floatfoot{\footnotesize \textbf{Note:} The probability (unconditional and conditional) of remaining as top-supplier are computed through a regression exercise. The regression takes the form of $\mathcal{P}_{i,j,t+s} = \beta_0 + \epsilon_{i,j,t+1}$ is estimated with, where $\mathcal{P}_{i,j,t+s}$ is an indicator function which takes the value of 0 if supplier $j$ is no longer firms top-supplier in period $t+s$ and takes the value of 1 if the supplier $j$ remains as top supplier in period $t+s$. Parameter $\beta_0$ is reported in the figures. The difference across Figures \ref{fig:persistence_unconditional} and \ref{fig:persistence_conditional} is that for the latter I condition on the link between firm $i$ and supplier $j$ being active in period $t+s$. For additional details and results on these regressions see Table \ref{tab:SF_persistence} in Appendix \ref{sec:appendix_stylized_facts}.}
\end{figure}
Figure \ref{fig:persistence_top_supplier} presents the probability that firm $i$'s supplier $j$ is her top-supplier in period $t+s$ conditional on being top-supplier in period $t$. On the left panel, Figure \ref{fig:persistence_unconditional} presents this probability for periods $t+s$ for $s=\{1,2,3\}$. The probability of supplier $j$ remaining as a firm's top-supplier is 40\% one year ahead, close to 30\% two periods ahead, and 25\% three periods ahead. The right panel, computes this probability conditional on the linkage with supplier $j$ being active in periods $t+s$. The conditional probability of supplier $j$ remaining as a firm's top-supplier is 70\% one year ahead, close to 68\% two periods ahead, and 65\% three periods ahead. The fact that linkages between a domestic firm and her top-supplier are more persistent than with other linkages are in line with the theory of search and informational frictions in the international markets for foreign suppliers.

In summary, this section of the paper presented evidence of search and informational frictions in the market for foreign suppliers. I show that a large share of domestic firms establish new linkages with foreign suppliers every year. Furthermore, I provide evidence that while linkages have a low probability of remaining active in subsequent periods, top-supplier linkages are relatively more persistent and new-linkages are less likely to survive. These facts combined suggest that domestic firms' network of foreign suppliers exhibit a high-turnover, a stylized fact previous literature has associated with search and informational frictions.

%%%%%%%%%%%%%%%%%%%%%%%%%%%%%%%%%%%%%%%%%%%%%%%%%%%%%%%%%%%%%%%%%%%%%%%%%%%%%%%%%%%%%%%%%%%%%%%%%%%%%%%%%%%%%%%%%%%%%%%%%%%%%%%%%%%%%%%%%%%
\section{Theoretical Framework} \label{sec:model}

In this section of the paper I develop a theoretical framework that guides the empirical work presented in Section \ref{sec:supplier_shocks}. The theoretical framework is one of heterogeneous firms which face search frictions on the market for foreign suppliers of inputs and endogenously choose whether to export or not. I separate firms' intertemporal profit maximization problem into a ``\textit{Static Problem}'' and a ``\textit{Search Problem}'', described in Sections \ref{subsec:model_static_stage} and \ref{subsec:model_search_stage} respectively. Section \ref{subsec:model_numerical} describes the model features and predictions through calibration and numerical exercises.

Firm $i$ maximizes the discounted present value of profits
\begin{align} \label{eq:discounted_profits}
    \Pi = \sum^{\infty}_{t=0} \beta^t \pi \left(z_i,C_{i,t}\right)
\end{align}
where $z_i$ and $C_{i,t}$ are firms idiosyncratic productivity and marginal cost of production respective described in further detail in the next section, and $\beta$ is a discount factor. The timing of the model is the following: (1) period $t$ starts with firm $i$ observing their idiosyncratic productivity process $z_i$ and set of foreign suppliers indexed by $k$ for $k=1,...,K$, (2) given this set of variables and the model's parameters a firm chooses which markets to serve, produces and sells, (3) firm decides whether to search or not for a new foreign supplier, (4) if the firm searches, it pays the fixed cost and observes the efficiency of the new supplier.

%%%%%%%%%%%%%
\subsection{Static Problem} \label{subsec:model_static_stage}

Firms are heterogeneous in their productivity which allows them to produce using a lower amount of inputs. The firm's static problem is comprised of production and selling to a domestic market and (potentially) to a foreign market. 

First, I describe the domestic and foreign market demand schedules, which I denote with the letters $d$ and $f$, respectively. I assume that in each country there is a representative consumer with CES preferences such that a domestic firm $i$ in country $j$ faces demand schedule
\begin{align}
    x_{i,j} = \left(\frac{p_{i,j} }{P_j}\right)^{-\rho} Y_j
\end{align}
where $x_{i,j}$ and $p_{i,j}$ are quantities sold and price charged by firm $i$ in market $j$, $P_j$ and $Y_j$ are the aggregate price index and the demand of market $j$, and $\rho$ is a parameter that governs the demand's elasticity of substitution. I assume that in order to actively sell in foreign market $f$, firm $i$ must pay a fixed cost $F^{e}$ every period. 

Second, I describe firm $i$'s production function and the role of imported inputs in it. I assume that production for different markets is carried out in separate production lines. I also assume that production functions are comprised of two CES tiers. The first tier is an aggregator between labor and a bundle of foreign inputs
\begin{align} \label{eq:production_first_tier}
    x_{i,j} \left(L_{i,j}, M_{i,j}\right) = z_i \left[\alpha L^{\theta}_{i,j} + \left(1-\alpha\right) M^{\theta}_{i,j}\right]^{\frac{1}{\theta}}
\end{align}
where $z_i$ represents firm $i$'s idiosyncratic productivity, $L_{i,j}$ and $M_{i,j}$ are the amount of labor hired and the bundle of imported inputs used by firm $i$ in production line $j$, and $\theta\in\left(0,1\right)$ governs the elasticity of substitution between production inputs. Cost minimization implies that optimal input demands and the unit cost of the first CES tier for a given production level $\bar{x}$ are given by
\begin{align*}
     L_{i,j} &= \left(\frac{\alpha C_{i,j} }{w}\right)^{\frac{1}{1-\theta}} \bar{x} \\
     M_{i,j} &= \left(\frac{\left(1-\alpha\right) C_{i,j} }{P^{M}_{i,j}}\right)^{\frac{1}{1-\theta}} \bar{x} \\
     C_{i,j} \left(w, p^M_{i,j} \right) &= \left(\alpha^{\frac{1}{1-\theta}} w^{\frac{\theta}{\theta-1}} + \left(1-\alpha\right)^{\frac{1}{1-\theta}} \left(p^M_{i,j}\right)^{\frac{\theta}{\theta-1}}\right)^{\frac{\theta-1}{\theta}}
\end{align*}
where $w$ is the wage rate and $p^M_{i,j}$ is the cost of the bundle of imported inputs in production line $j$. Thus, firm $i$'s marginal cost of production in line $j$ is given by $ C_{i,j}/z_i$.

The bundle of imported inputs is itself a CES aggregator of different varieties of imported inputs. In particular, I assume that $M_{i,j}$ is of the form
\begin{align} \label{eq:production_second_tier}
    M_{i,j} = \left[\sum^{K}_{k=1} m^{\varphi}_{i,j,k} \right]^{1/\varphi}
\end{align}
\begin{align*}
    p^M_{i,j} &= \left[\sum^{K}_{k=1} p^{\frac{\varphi}{\varphi-1}}_{i,j,k} \right]^{\frac{\varphi-1}{\varphi}} \\
    m_{i,j,k} &=  \left(\frac{p^M_{i,j}}{p_{i,j,k}}\right)^{\frac{1}{1-\varphi}} \bar{M} \quad \text{for } k=1,\ldots,K
\end{align*}
Note that given the independence of production lines and the constant returns to scale of the production technology, firm $i$ uses the same set of foreign suppliers across production lines. Thus, I drop the sub-index $j$ for the bundle of imported inputs and for individual varieties. 
\noindent
Firm $i$'s production line $j$'s profit maximization problem can be stated as
\begin{equation*}
\begin{aligned}
\max_{x_{i,j},p_{i,j}} \pi_{i,j} \left(z_i,\tau_j\right) = \quad & \left(p_{i,j} - \frac{\tau_j}{z_{i}} C_{i} \right) x_{i,j} - e_jF^{e} \mathbbm{1}\left[x_{i,j}>0\right] \\ %- f |\Omega_{i,j}|^{\mu} \\
\textrm{s.t.} \quad & x_{i,j} = p^{-\rho}_{i,j} Y_j P^{\rho}_j \\
  &  C_{i} \left(w, p^M_{i} \right) = \left(\alpha^{\frac{1}{1-\theta}} w^{\frac{\theta}{\theta-1}} + \left(1-\alpha\right)^{\frac{1}{1-\theta}} \left(p^M_{i}\right)^{\frac{\theta}{\theta-1}}\right)^{\frac{\theta-1}{\theta}} \\
  &  p^M_{i} = \left[\sum^{K}_{k=1} p^{\frac{\varphi}{\varphi-1}}_{i,k} \right]^{\frac{\varphi-1}{\varphi}}
\end{aligned}
\end{equation*}
where $\tau_j$ is market $j$'s iceberg cost and $e_j$ is binary variable which takes the value of 1 if $j$ is equal to $f$ and zero otherwise. I assume that production for the domestic market implies no iceberg costs, i.e. $\tau_d =0$, while exporting implies positive iceberg costs, i.e., $\tau_f>0$.

\noindent
\textbf{Proposition 1.} For a given wage rate $w$, parameters $\{\alpha, \rho, \theta, \varphi, P_f, Y_f, \tau_f \}$ and a given set of imported varieties prices $\{p_{i,k}\}^{K}_{k=1}$, firm $i$ exports if the idiosyncratic productivity $z_i$ is greater than $\bar{z}$, given by
\begin{align} \label{eq:export_threshold}
    \bar{z} = \frac{\tau_f C_{i}}{\rho-1} \left[ \rho^{\rho} \frac{w F^e}{P^{\rho}_f Y_f} \right]^{\frac{1}{\rho-1}}
\end{align}
If $z_i \geq \bar{z}$, firm $i$'s exported quantities are given by
\begin{align} \label{eq:optimal_quantities_exported}
    x_{i,f} =  \left(\frac{\tau_f}{z_i}C_{i,j}\right)^{-\rho} P^{\rho}_f \left( \frac{\rho}{\rho-1}\right)^{-\rho} Y_f
\end{align}
\textit{See proof in Appendix \ref{sec:appendix_proofs}.} 

\begin{comment}
\noindent
\textbf{Corollary 1.} For a given wage rate $w$, parameters $\{\alpha, \rho, P_f, Y_f, \tau_f \}$, a given set of imported varieties prices $\{p_{i,k}\}^{K}_{k=1}$ and firm productivity $z_i$, the impact of an increase in the price of an individual foreign variety $k$ on the exporting threshold $\bar{z}$ and exported quantities is decreasing in $\theta$.
\noindent 
\textit{See proof in Appendix \ref{sec:appendix_proofs}.} 
\end{comment}

Proposition 1 shows that only a subset of highly productive firms export, in terms of being able to produce at low marginal costs. Furthermore, conditional on exporting, lower marginal costs lead to higher exported quantities. This is in line with the canonical model of \cite{melitz2003impact}. However, under the current framework, firms can produce at low marginal costs for two different reasons: (i) a high idiosyncratic productivity process $z_i$, (ii) a low marginal input cost $C_i$. For a given wage, firms differ in the cost of their input bundle by the heterogeneity of their set of imported varieties. Firms with access to a set of imported inputs with low prices $p_{k}$ are able to produce at lower marginal costs than firms with a set of imported inputs with high prices $\tilde{p}_{k}$. Furthermore, as stressed by \cite{gopinath2014trade}, not only prices $\tilde{p}_{k}$ matter, but also the cardinality $K$ of the set, as a higher number of foreign suppliers decreases the overall cost of the imported input bundle.\footnote{This is because the CES production function in Equation \ref{eq:production_second_tier} exhibits love for varieties.} Consequently, even conditional on idiosyncratic productivity $z_i$, firms can have different export status (exporting or not) and/or different exported quantities, driven by differences in their set of foreign suppliers.

%%%%%%%%%%%%%
\subsection{Search Problem \& Supplier Set Dynamics} \label{subsec:model_search_stage}

Next, I turn to describing firm $i$'s search problem and the dynamics of her set of suppliers. The firm's search problem determines the number of foreign suppliers $K$ used in production. However the market for these suppliers exhibit search and matching frictions which take the form of searching implying the payment of a fixed cost and leading to a random match with foreign suppliers which are heterogeneous in their efficiency. 

Foreign suppliers differ in their prices reflecting differences in efficiency. I denote with $G\left(p_k\right)$ and $g\left(p_k\right)$ the cumulative and probability density functions of $p_k$ which I assume to be continuously differentiable in $p_k$ between bounds $\{\underline{p},\bar{p}\}$.\footnote{In Section \ref{subsec:model_numerical} where I carry out numerical exercises I assume that the distribution of $p_k$ is uniform.} Additionally, I assume that foreign supplier efficiencies are independent of each other. 

Every period a firm can choose whether to search for a new foreign supplier or not. I assume that a domestic firm can only search for one new supplier every period. This search involves the payment of a fixed cost $F^S$ which follows the form
\begin{equation} \label{eq:fixed_cost_function}
    F^S\left(K\right) = f^S \times \left(K-1\right)^{\mu}
\end{equation}
where as before $K$ is the cardinality of firms' set of foreign suppliers, $f^S$ is a constant term and $\mu>0$ is a parameter that governs the convexity of function $F^S(K)$. Note that this functional form implies that searching for the first foreign supplier has no cost. This is line with \cite{gopinath2014trade} which show that aggregate import dynamics are not explained by entry and exit of firms from import markets.\footnote{However, the functional form in Equation \ref{eq:fixed_cost_function} differs from that used by \cite{gopinath2014trade}. Instead of modelling firms' decision over a set of foreign varieties, in this paper I choose a functional form which explicitly depends on the cardinality of firms' set of suppliers. This is inspired by the stylized facts presented in Section \ref{subsec:stylized_facts}.} The increasing nature of this function could be explained by decreasing returns to scale in searching or increasing cost of maintaining already active foreign supplier linkages.

In order to determine whether a firm searches for a new supplier she computes the expected payoff of this search. Firm $i$'s expected payoff is the difference between next period's expected per period profits across all possible realizations of the foreign supplier $K+1$'s efficiency and its current profits given her current set of foreign suppliers $k=1,\ldots,K$ and its productivity level $z_i$. If I denote firm $i$'s current period profits $\pi \left(z_i,\{p_{i,k}\}^{K}_{k=1} \right)$ and next period profits after it searched and matched with supplier $s$ as $\pi \left(z_i,\{p_{i,k}\}^{K}_{k=1},p_{s} \right)$, then the expected payoff of searching is given by the difference between the mean profits across all possible realizations and the current period profits
\begin{align} \label{eq:search_expected_profits}
    \underbrace{\pi^S\left(z_i,\{p_{i,k}\}\right)}_{\text{Expected Payoff of Searching}} = \underbrace{\int^{\bar{p}}_{\underline{p}} \pi \left(z_i,\{p_{i,k}\}^{K}_{k=1},p_{s} \right) g\left(p_s\right) ds}_{\text{Expected next period's profits}} - \underbrace{\pi \left(z_i,\{p_{i,k}\}^{K}_{k=1} \right)}_{\text{Current profits}}
\end{align}
Given firms' objective function in Equation \ref{eq:discounted_profits}, firm $i$ with productivity $z_i$ and set of linkages with foreign suppliers $\{p_{i,k}\}^{K}_{k=1}$ searches if
\begin{align} \label{eq:search_condition}
    \frac{\beta}{1-\beta} \times \pi^S\left(z_i,\{p_{i,k}\}^K_{k=1}\right) \geq F^{S}\left(K\right)
\end{align}
The condition above states that a firm searches for a new foreign supplier if the expected discounted increase in profits from searching and purchasing goods from a new supplier is greater than the fixed cost involved in the search. Note that the greater the fixed cost of searching $f^S$ and the greater the convexity parameter $\mu$, the greater $F^{S}(K)$, the less likely it is for a firm to search for a new foreign supplier. The term $\beta/\left(1-\beta\right)$ reflect the fact that firm $i$ decides in period $t$ whether to search or not, and if she does, the new foreign supplier is added to her set in period $t+1$. Furthermore, the new expanded foreign supplier becomes her benchmark for comparison for period $t+1$ onward. 

%%%%%%%%%%%%%
\subsection{Numerical Exercises} \label{subsec:model_numerical}

Next, I shed light on the model's features through numerical exercises. In particular, I show how search and informational frictions and imperfect substitutability across foreign suppliers matter for the transmission of a supplier-specific shock. I begin by describing the model's parametrization, primarily carried out using values used by previous literature and calibrating a subset of parameters. The main goal of this exercise is not to match all of the features of firms' network of foreign suppliers presented in Sections \ref{subsec:stylized_facts} and \ref{subsec:search_evidence}, but to provide a benchmark parametrization to study the implications of search and matching frictions and the impact of a foreign supplier-supply shock. To this end, I simulate a large number of firms show how aggregate moments and the transmission of supplier-specific shocks as a function of parameters $\{f^s,\mu,\bar{p}\}$.  

\noindent
\textbf{Parametrization \& calibration.} First, I describe the subset of parameter values taken from previous literature. The demand CES elasticity $\rho$ is set equal to $5$, in line with previous literature \cite{kohn2016financial}. The weight on labor on the first tier of the production function in Equation \ref{eq:production_first_tier} is set equal to $2/3$ in line with Argentina's input-output table from 1996 and \cite{gopinath2014trade}.
\begin{table}[ht]
    \centering
    \begin{tabular}{c|c|c}
    Parameter  & Description & Value  \\ \hline \hline
    $\rho$     & CES Demand elasticity parameter & 5 \\
    $\alpha$   & Labor Weight First-Tier   & 2/3 \\ 
    $\theta$   & CES Parameter First-Tier  & 0.50 \\
    $\tau_f$   & Iceberg Cost              & 1.50 \\
    $w$        & Wage rate                 & 1 \\ 
    $\varphi$  & Substitution across varieties of imported inputs & 0.75 \\
    \end{tabular}
    \caption{Model Parametrization}
    \label{tab:parametrization}
\end{table}
The CES parameter which governs the elasticity of substitution between the bundle of imported inputs and labor is $\theta=0.50$, which implies an elasticity of substitution of 2. This value is between the unitary elasticity of substitution between labor and inputs and the elasticity of substitution between domestic and foreign inputs calibrated in \cite{gopinath2014trade}. Thus, the present parametrization implies that the production function in \ref{eq:production_first_tier} represents a technology which bundles domestic and foreign inputs. Iceberg costs are set equal to $\tau_f = 1.50$, in line with \cite{kohn2016financial}. Setting the wage rate at 1 is a normalization. Finally, I set parameter $\varphi$, which governs the elasticity of substitution between varieties of imported intermediate inputs equal to 0.75 from \cite{gopinath2014trade}. This value implies an elasticity of substitution of 4 across intermediate input produced by different suppliers.\footnote{\cite{gopinath2014trade} showed that a value of $\varphi = 0.75$ is consistent with other features of the data for the same time period.}

Second, I turn to describing the calibration of the rest of the parameters in the model. I assume that firms' idiosyncratic productivity follows a log-normal distribution with mean $\mu_z = 0.5$ and variance $\sigma_z$. The choice of $\mu_z=0.5$ is a normalization, as only the ratio $\mu_z/\sigma_z$ can be identified.\footnote{See \cite{fanelli2015export} for a discussion on this identification topic.} I also normalize $\underline{p} = 0.5$. Thus, I have to assign numerical values for five parameters: (i) parameter set $\Theta = \{f^s,\mu,\bar{p}\}$, and (ii) $\{\sigma_z, F^{e}\}$. I calibrate parameter values in order to minimize the difference between a set of model and data moments. This exercise seeks to match (i) the mean number of foreign suppliers, (ii) the median number of foreign suppliers, (iii) the mean share of firms' imported value explained by her top-supplier, (iv) the distribution of imported value per firm and (v) the share of exporting firms. To construct the model counterparts of these moments from the data I simulate 5,000 firms by taking draws of the distribution of productivity $z_i$. I simulate firms' search choice for foreign suppliers until their supplier set converges and firms have no incentive to continue searching. Once firms' stop searching for new suppliers I compute the model's simulated moments.\footnote{For additional details on the calibration exercise see Appendix \ref{sec:appendix_model_calibration_fit}.}

The results of the calibration exercise, parameter values and features of the data can be found in Table \ref{tab:calibration}.
\begin{table}[ht]
    \centering
    \caption{Model Calibration}
    \label{tab:calibration}
    \small 
    \begin{tabular}{c c || l c c}
       Parameter & Value & Target Moment & Data & Model \\ \hline \hline
       $f^s$      & 0.0046 & Mean Number of Suppliers    & 6.11 & 6.0668\\
       $\mu$      & 0.6079 & Median Number of Suppliers  & 2.00 & 2.00 \\
       $\bar{p}$  & 4.4974 & Mean Share explained by Top-Supplier & 0.746 & 0.6462   \\
%       $\varphi$  & 0.750 & Median Share explained by Top-Supplier  & 0.832 & 0.9576 \\
       $\sigma_z$ & 0.0267 & Distribution of Imported Value per Firm & \multicolumn{2}{c}{See Figure \ref{fig:cum_share_impo}}  \\
       $F^{e}$    & 0.0049 & Share of Exporting Firms & 0.10 & 0.102 \\
    \end{tabular}
\end{table}
Overall, the calibration of the model does a good job in matching the moments of the data. The model does a good job in matching statistics over the number of foreign suppliers, but is less precise when matching the share of total imported value explained by firms' top supplier. The model matches the share of exporting firms. Finally, Figure \ref{fig:cum_share_impo} in Appendix \ref{sec:appendix_model_calibration_fit} presents the model's implied distribution of imported value per firm and its data counterpart, showing an accurate fit.\footnote{Appendix \ref{sec:appendix_model_calibration_fit} presents additional evidence on the model calibrations' fit to the data.}

\noindent
\textbf{Impact of a Supplier-Supply-Shock.} I turn to computing  the impact on firms imported and exported quantities. In particular, I compute the impact of a 15\% increase in the cost of imports by a firm's top-supplier.\footnote{The choice of a 15\% increase in the cost of firms' top supplier is motivated by the standard deviation of the empirical firm level Supplier-Supply-Shock constructed in Section \ref{subsec:supplier_supply_shocks}.}

\begin{figure}[ht]
    \caption{Model's Supplier Supply Shock}
    \label{fig:model_supplier_cost_shock}
     \centering
     \begin{subfigure}[b]{0.495\textwidth}
         \centering
         \includegraphics[width=\textwidth]{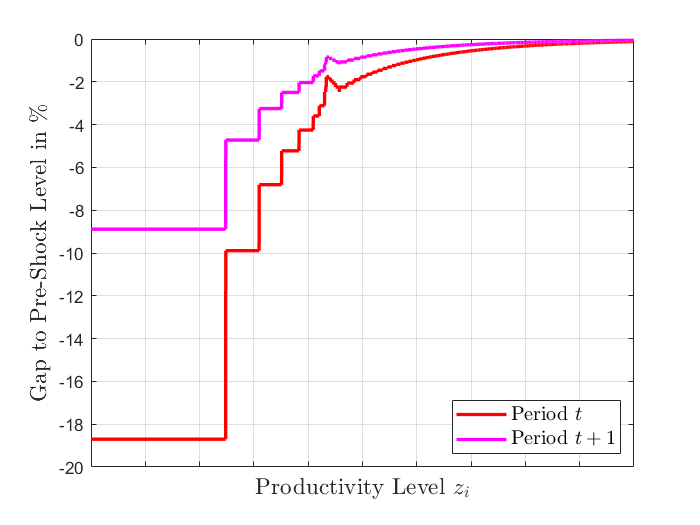}
         \caption{Imported Quantities}
         \label{fig:IRFs_top_imports}
     \end{subfigure}
     \hfill
     \begin{subfigure}[b]{0.495\textwidth}
         \centering
         \includegraphics[width=\textwidth]{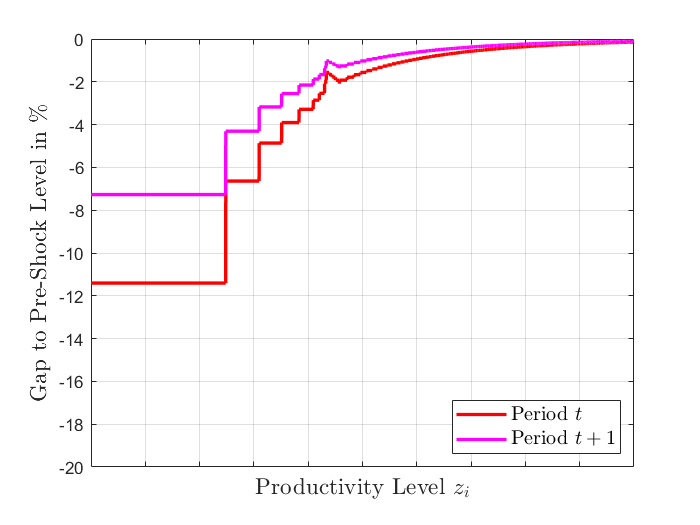}
         \caption{Exported Quantities}
         \label{fig:IRFs_top_exports}
     \end{subfigure}
%    \floatfoot{\footnotesize \textbf{Note:} }
\end{figure}

I start by computing the impact of this increase in the cost of firms' top-supplier under the parametrization described in Tables \ref{tab:parametrization} and \ref{tab:calibration}. Figure \ref{fig:model_supplier_cost_shock} presents the results of this numerical exercise for the period of the shock (solid red line), denoted period $t$ and the subsequent period, or period $t+1$ (solid magenta line).
On the left panel, Figure \ref{fig:IRFs_top_imports} shows the drop in imported quantities of a 15\% increase in the cost of firms' top supplier for different values of firm's idiosyncratic productivity level $z_i$. The first takeaway from this figure is that relatively less productive firms experience a larger drop in imported quantities of inputs than relatively more productive firms. This is due to the fact that more productive firms can afford to search for more suppliers and consequently, be able to immediately substitute away from the \textit{shocked} supplier (than less productive firms). The second takeaway is the jagged nature of heterogeneous impact across firms with different idiosyncratic productivity $z_i$. This is generated by the granularity of firms' supplier set. Relatively less productive firms have smaller supplier sets than more productive firms. Given that the median firm only as 2 suppliers, the change in magnitude of impact is more pronounced for relatively smaller firms as the gains from gaining suppliers is significantly high. The final takeaway is that the increase in the cost of firms' top suppliers leads firms into searching for new suppliers. This can be observed from the relatively lower impact in imported quantities in period $t+1$ (solid magenta line). As firms search for new suppliers, they can better substitute away from the \textit{shocked} supplier.  

On the right panel, Figure \ref{fig:IRFs_top_exports} presents the impact of a 15\% increase in the cost of firms' top supplier input on exported quantities. In line with the model's results for imported quantities, relatively smaller and less productive firms experience a larger drop in exported quantities than relative larger more productive firms. The magnitude of impact in exported quantities is lower than for imported quantities as firms can substitute away from the imported input bundle towards labor. However, the model calibration implies that the ratio of fall in exported quantities to the fall in imported quantities has a mean and median of 0.8716 and 0.8873 respectively. This is, the model predicts a strong complementarity between imported inputs and export performance.

\noindent
\textbf{Sensitivity Analysis.} Lastly, I show how aggregate moments and the impact of Supplier-Supply-Shock depends on parameters $\{f^s,\mu,\bar{p},\varphi\}$. Figure \ref{fig:Sensitivity_Analysis} in Appendix \ref{sec:appendix_model_calibration_fit} shows how the average number of foreign suppliers and the mean share of imported value explained by the top supplier depend on parameters $\{f^{s},\mu,\bar{p},\varphi\}$. As expected, the higher the fixed cost $f^{s}$ and convexity of the fixed cost function $\mu$ reduces the average number of foreign suppliers and increases the mean share of imported value explained by firms' top-supplier. Interestingly, as $\bar{p}$ increases, the average number of foreign suppliers and the mean share of imported value explained by the top-supplier also decline. This is because as $\bar{p}$ increases the mean and variance of the distribution of prices of foreign supplier's efficiency increases. By decreasing the average efficiency of foreign suppliers, firms' incentives to search for foreign suppliers are reduced. Lastly, an increase in the elasticity of substitution across intermediate input suppliers $\varphi$, firms average number of foreign suppliers and the . As the elasticity of substitution increases, the love for variety of production function in Equation \ref{eq:production_second_tier} drops. In the limit, as $\varphi \rightarrow 1$ and intermediate inputs become perfect substitutes, a firm only uses their most efficient foreign supplier. This leads to firms reducing their number of foreign suppliers and focusing most of their imported value on their most efficient supplier.

The second sensitivity analysis is to test how changes in parameters $\{f^{s},\mu,\bar{p},\varphi\}$ matter for the impact of a Supplier-Supply shock. Figure \ref{fig:IRFs_sensitivity} in Appendix \ref{sec:appendix_model_calibration_fit} presents the results of this exercise. Higher fixed costs of searching, $f^{s}$, and higher convexity parameter $\mu$ aggravate the impact of an increase of a Supplier Supply shock. However, the increase in impact is heterogeneous across firms. Firms which have only one foreign supplier under the benchmark parametrization are not affected by changes in $f^{s}$ and $\mu$. However, higher $f^{s}$ and $\mu$ shift the distribution of firms' cardinality of supplier set to the left, i.e., firms now have a lower number of foreign suppliers. For these firms, the impact of a Supplier Supply shock is greater. An increase in the elasticity of substitution between intermediate inputs, $\varphi$, also aggravates the impact of a Supplier Supply shock. As stressed above, higher $\varphi$ imply that firms have less incentives to search for foreign suppliers as their production function loses gains driven by love-for-variety. Consequently, a reduction in the number of foreign suppliers leaves firms more exposed to an idiosyncratic supplier shocks.

In summary, this section presented a theoretical framework to study how a supplier supply cost shock impacts firms' export performance in the presence of search and matching frictions in international markets for suppliers. The model was parametrized and calibrated to match several features the stylized facts presented in Section \ref{sec:data_facts} and other moments of the data. Simulating a supplier cost shock, the model predicts that a higher cost of imported inputs should be reflected in lower exported quantities, i.e., a high degree of import-export complementarity. Furthermore, the model predicts that smaller and less productive firms should experience a larger drop in both imported and exported quantities, given that these firms have less diversified and smaller scale networks of foreign suppliers. In the next section, I construct firm-level supplier supply shocks and test these predictions in the data.

%%%%%%%%%%%%%%%%%%%%%%%%%%%%%%%%%%%%%%%%%%%%%%%%%%%%%%%%%%%%%%%%%
\section{Supplier - Supply Shocks \& Export Performance} \label{sec:supplier_shocks}

In this section I present the main empirical regression results of the paper. Motivated by the search and matching frictions introduced in the model in Section \ref{sec:model}, I construct a firm level measure of supplier-supply shocks and estimate their impact on import and export performance. Section \ref{subsec:supplier_supply_shocks} describes the proposed identification strategy based on exploiting the supplier dimension of the dataset used in Section \ref{sec:data_facts} to identify the two key novel stylized facts. Sections \ref{subsec:import_performance} and \ref{subsec:export_performance} estimate the impact of the identified supplier-supply shocks on imported and exported quantities respectively. 

%%%%%%%%%%%
\subsection{Identifying Supplier - Supply Shocks} \label{subsec:supplier_supply_shocks}

I turn to identifying firm-specific supplier-supply import cost shocks. The econometric approach presented in this section exploits the supplier dimension of the dataset to disentangle firm-specific demand shocks from supplier supply shocks. 

The dataset allows me to observe import transactions between domestic firm $i$ and foreign supplier $s$ of product $p$ from source country $d$. I label this detailed level of transaction as an import instance. For ease of notation, I denote an import instance at the $\{p,d,s\}$ level as $k$. Then, log price changes in import prices at the instance $k$ level can be expressed as
\begin{align} \label{eq:change_prices}
    %\Delta  \tilde{P}_{i,k,t} = \frac{ P_{i,k,t} - P_{i,k,t-1}}{0.5 \times \left(P_{i,k,t} + P_{i,k,t-1} \right)}
    \Delta  \ln P_{i,k,t} = \ln P_{i,k,t} - \ln P_{i,k,t-1}
\end{align}
\textit{A priori}, a measure of firm $i$ changes in import prices can be expressed as
\begin{equation} \label{eq:first_specification}
    %\Delta P_{i,t} = \sum_{k=1} s_{i,k,t} \times \Delta  \tilde{P}_{i,k,t}
    \Delta \ln P_{i,t} = \sum_{k=1} s_{i,k,t} \times \Delta  \ln P_{i,k,t}
\end{equation}
where $s_{i,k,t}$ represents the share of firm $i$'s total value imported explained by instance level $k$ in period $t$. In principal, changes in instance-level prices $\Delta \ln P_{i,k,t}$ can be explained by several different sources of variation. On the one hand, firm $i$ may be subject to time specific demand shocks which increase transportation costs unilaterally across import instances. On the other hand, product specific shocks (such as an oil shocks) or destination specific shocks (such as a country suffering a natural disaster) may explain variations on Argentinean firms' import prices. While these may be interesting sources of import price shocks, in this paper I focus on identifying supplier-supply shocks. 

Next, I turn to describing the construction of firm $i$ specific supplier-supply shocks and the concerns that arise when trying to identify them. The first empirical concern is that the weighted price change measure in Equation \ref{eq:first_specification} may be contaminated by firm-specific demand shocks. In other words, it may be the case that changes in productivity or firm-specific transportation costs affect all import prices of firm $i$. The second concern is that the shares $s_{i,k,t}$ in Equation \ref{eq:first_specification} are endogenous variables. This is, after a positive import price increase $\Delta P_{i,k,t}$, firm $i$ may reduce the imported value of instance $k$, or even drop the instance all together.

First, I seek to extract a supplier specific import price shock from changes in import prices. A first specification that cleanses import price changes $\Delta \ln P_{i,p,d,s,t}$ from firm specific demand shocks is given by the following empirical regression
\begin{align} \label{eq:price_reg}
    \Delta \ln P_{i,p,d,s,t} = \beta_{i,t} + \epsilon_{i,p,d,s,t}
\end{align}
where $\beta_{i,t}$ represent firm-time fixed effects, and $\epsilon_{i,p,d,s,t}$ is an orthogonal error term. The error term $\epsilon_{i,p,d,s,t}$ represents a mix of all possible combinations of supply import price shocks. Thus, to cleanse import prices from firm specific demand shocks and dissect supplier-specific supply shocks I estimate the following empirical regression
\begin{align} \label{eq:price_reg_supplier}
    \Delta  \tilde{P}_{i,p,d,s,t} = \tilde{\beta}_{i,t} + \gamma_{s,t} + \tilde{\epsilon}_{i,p,d,s,t}
\end{align}
where $\tilde{\beta}_{i,t}$ represents firm-time fixed effects, $\gamma_{s,t}$ represents supplier-time fixed effects, and $\tilde{\epsilon}_{i,p,d,s,t}$ is an error term orthogonal to firm demand shocks and supplier-time shocks. Thus, $\gamma_{s,t}$ captures the variation in changes in Argentinean firms' import prices that is explained by a common supplier-time specific component. This variation is identified as foreign suppliers sell multiple products, from different source countries to different Argentinean customer firms.\footnote{For more details on the distribution of foreign-suppliers' number of products exported, source country origins, export instances and Argentinean customers see Appendix \ref{sec:appendix_features_data}.} Thus, under the current set of assumptions $\gamma_{s,t}$ captures a supplier-specific supply variation of import prices.

Second, I address the concerns over the shares $s_{i,k,t}$ in Equation \ref{eq:first_specification}. First, note that domestic firms vary in their exposure to different foreign suppliers which serve Argentinean firms. Thus, the shares $s_{i,k,t}$ provide a significant source of heterogeneity across firms. This heterogeneity arises from the high dimensional space of foreign suppliers the Argentinean firms are exposed to. For instance, Argentinean firms import from 220 different countries, 4,961 different products and from 512,915 different suppliers. Firms have imported from 3,470,818 distinct combinations of these three variables.\footnote{Out of 559,805,689,300 possible combinations if the number of foreign supplier in the sample reflect the actual total number of international or world suppliers.} This high degree of heterogeneity in sourcing strategies makes it unlikely that any two Argentinean firms are exposed exactly the same to the same market. A second source of heterogeneity across firms is each firms' link intensity or relevance of trade flow. In other words, firms $X$ and $Y$ may import the same product from the same source country and even from the same supplier, but this trade flow could explain $10\%$ of firm $X$'s total imports basket but represent $90\%$ of firm $Y$'s import bundle. 

Note that even if the fixed effects $\gamma_{s,t}$ represent an exogenous measure of supplier-specific shocks to Argentinean firms, these firms choose their network of foreign suppliers endogenously. As argued above, firms can react to import prices changes in import instance $k$ by varying the intensity of the trade flow or disrupting it completely. Thus, time $t$ shares, i.e. $s_{i,k,t}$ from Equation \ref{eq:first_specification}, already incorporate this endogenous firm reaction. To control for this endogenous firm behavior I use time $t-1$ shares based on the intuition presented in \cite{ad2019shift}'s shift-share design. The reasoning behind this choice is that Argentinean firms choose which instances $k$ to import from in period $t-1$, establish trade linkages, and are subject to the multiple possible supply shocks arising from them in period $t$. This is line with the search and matching frictions introduced to the theoretical framework constructed in Section \ref{sec:model}.

Now, I have the two necessary elements to construct firm $i$'s supplier supply import price shock. The following expression combines the supplier-time fixed effects $\gamma_{s,t}$ recovered from estimating Equation \ref{eq:price_reg_supplier}, and weights $\omega_{i,s,t-1}$ which represents the share of firm $i$'s total imports explained by supplier $s$ in period $t-1$
\begin{align}
    \text{Supplier-Supply-Shock}_{i,t} = \sum_s \omega_{i,s,t-1} \times \gamma_{s,t}
\end{align}
This shock captures firm $i$'s increase in the cost of imported inputs explained by the suppliers-linkages in place in period $t-1$. Note that if firm $j$ in Argentina also purchased from supplier $s$ in period $t-1$, then she is also exposed to the supplier-specific shock $\gamma_{s,t}$, with the exposure defined by $\omega_{j,s,t-1}$. 

The impact this shock has on firms' imported quantities and overall performance depends crucially on the degree of substitutability across suppliers. On the one hand, if firm $i$ observes a price increase from supplier $s$ and can immediately substitute and purchase from supplier $\tilde{s}$, then the $\text{Supplier-Supply-Shock}_{i,t}$ should not have a negative impact on its performance. On the other hand, if firm $i$ can not substitute across foreign suppliers, then she will be negatively impacted. 

%%%%%%%%%%%
\subsection{Impact on Imported Quantities} \label{subsec:import_performance}

This section of the paper describes the impact of the supplier-supply shocks on firms' imported quantities. \textit{A priori}, one would expect that higher import costs would lead to a reduction in imported quantities. This is straightforward for the quantities demanded from the supplier who increased prices. In addition, as the firm-level import bundle cost increases, she will shift from imported products towards labor or domestically produced inputs. As stressed above in Section \ref{subsec:supplier_supply_shocks}, and by the predictions of the model presented in Section \ref{sec:model} the magnitude of impact depends on the size of firms' network of foreign suppliers and how substitutable intermediate goods are with each other.

To measure the impact of a supplier-supply import cost shock I estimate the following regression specification
\begin{align} \label{eq:regression_imports_supplier}
    \ln QI_{i,p,d,s,t} = \beta^{I} \text{Supplier-Supply-Shock}_{i,t} + \gamma_{i,p,d,s} + \gamma_t + \Gamma_{i,p,d,t} + e_{i,p,d,s,t}
\end{align}
where $\ln QI_{i,p,d,s,t}$ represents log imported quantities by firm $i$ of product $p$ from source country $d$ and supplier $s$ in year $t$, $\gamma_{i,p,d,s}$ is a firm-product-source-supplier fixed effect which captures invariant features of the import flow of firm $i$ of product $p$ from source country $d$ and supplier $s$ (such as underlying overall quality of the match), $\gamma_t$ is a year fixed effect which captures aggregate shocks that affect all Argentinean firms equally (such as a wage shock), and $\Gamma_{i,p,d,t}$ represents additional control variables. In particular, I control for lagged values of log imported quantities, lagged values of the supplier-supply shock, current and lagged values of log of the bilateral real exchange rate between Argentina and source country $d$, and the log of total value imported by firm $i$ in period $t-1$.

Table \ref{tab:import_instances_supplier} presents estimates of $\beta^{I}$ for several specifications of Equation \ref{eq:regression_imports_supplier}. Across specifications, a positive supplier-supply shock leads to lower quantities imported at the $\{i,p,d,s\}$ level. 
\begin{table}[ht]
    \centering
    \caption{Impact of a Supplier-Supply Shock on Imported Quantities}
    \label{tab:import_instances_supplier}
    \footnotesize
\begin{tabular}{lcccc} 
 & \multicolumn{4}{c}{Imported Quantities - $\ln Q_{i,p,d,s,t}$} \\
 & (1) & (2) & (3) & (4) \\ \hline \hline
\\
$\text{Supplier-Supply-Shock}_{i,t}$ & -0.0150** & -0.0470*** & -0.0421*** & -0.0410*** \\
 & (0.00711) & (0.0122) & (0.0100) & (0.00950) \\
 &  &  &  &  \\
$\ln Q_{i,p,d}$ lags & NO & YES & YES & YES \\
$\text{Supplier-Supply-Shock}_{i}$ lags & NO & YES & YES & YES \\
$\ln$ RER$_{d,t}$ \& lags & NO & NO & YES & YES \\
$\ln$ Total imported value & NO & NO & NO & YES \\  
 &  &  &  &  \\
Observations & 2,098,170 & 650,785 & 541,475 & 541,475 \\ \hline \hline
% R-squared & 0.914 & 0.925 & 0.927 & 0.927 \\ \hline
\multicolumn{5}{c}{ Two-way clustered standard errors in parentheses (Firm-Source country level)} \\
\multicolumn{5}{c}{ *** p$<$0.01, ** p$<$0.05, * p$<$0.1} \\
\end{tabular}
\floatfoot{\footnotesize \textbf{Note:} Every specification in the table controls for $\gamma_{i,p,d,s}$ and $\gamma_t$ fixed effects. The specifications which control for lagged values of either the explained and/or the explanatory variables do so by controlling for the $t-1$ and $t-2$ values of the variables. }
\end{table}
To have a sense of magnitude of this shock. the estimated coefficients suggest that a one-standard deviation of this shock (16.882\% across the sample) leads to a drop in imported quantities between 0.25\% and 0.79\%.  These results suggest that a firm level supplier-supply shock significantly reduction in imported quantities, in line with the theoretical framework's predictions in Figure \ref{fig:IRFs_top_imports}. 

The results presented above show that a firm level supplier-supply shocks lead to lower quantities imported at the instance level $\{i,p,d,s\}$. As a robustness check, I estimate the following regression at the $\{i,p,d\}$ level. 
\begin{align} \label{eq:regression_imports}
    \ln QI_{i,p,d,t} = \tilde{\beta}^{I} \text{Supplier-Supply-Shock}_{i,t} + \gamma_{i,p,d} + \gamma_t + \Gamma_{i,p,d,t} + e_{i,p,d,t}
\end{align}
where $\ln QI_{i,p,d,t}$ represents log imported quantities by firm $i$ of product $p$ from source country $d$ in year $t$. Table \ref{tab:import_instances} presents estimates of parameter $\tilde{\beta}^{I}$ for specifications with different set of controls. 
\begin{table}[ht]
    \centering
    \caption{Impact of a Supplier-Supply Shock on Imported Quantities \\ \small Robustness Check}
    \label{tab:import_instances}
    \footnotesize
\begin{tabular}{lcccc} 
 & \multicolumn{4}{c}{Imported Quantities - $\ln Q_{i,p,d,t}$} \\
 & (1) & (2) & (3) & (4) \\ \hline \hline
\\
$\text{Supplier-Supply-Shock}_{i,t}$ & -0.0185*** & -0.0366*** & -0.0328*** & -0.0325*** \\
 & (0.00699) & (0.00987) & (0.00780) & (0.00769) \\
 &  &  &  &  \\
$\ln Q_{i,p,d}$ lags & NO & YES & YES & YES \\
$\text{Supplier-Supply-Shock}_{i}$ lags & NO & YES & YES & YES \\
$\ln$ RER$_{d,t}$ \& lags & NO & NO & YES & YES \\
$\ln$ Total imported value & NO & NO & NO & YES \\  
 &  &  &  &  \\
Observations & 1,971,366 & 836,056 & 699,143 & 699,143 \\ \hline \hline
% R-squared & 0.914 & 0.925 & 0.927 & 0.927 \\ \hline
\multicolumn{5}{c}{ Two-way clustered standard errors in parentheses (Firm-Source country level)} \\
\multicolumn{5}{c}{ *** p$<$0.01, ** p$<$0.05, * p$<$0.1} \\
\end{tabular}
\floatfoot{\footnotesize \textbf{Note:} Every specification in the table controls for $\gamma_{i,p,d}$ and $\gamma_t$ fixed effects. The specifications which control for lagged values of either the explained and/or the explanatory variables do so by controlling for the $t-1$ and $t-2$ values of the variables. }
\end{table}
In line with the results presented above, a positive supplier-supply shocks also leads to a drop in imported quantities at the $\{i,p,d\}$ level. 

Lastly, I test whether relatively smaller firms are hit harder by supplier-supply import cost shocks than relatively larger firms. Section \ref{sec:data_facts} shows that relatively smaller firms, in terms of employment, have a lower amount of foreign suppliers. Thus, if one of those suppliers increased the price of its imported input, \textit{a priori}, a smaller firm should be less able to substitute to another supplier than relatively larger firms linked with a higher number of suppliers. If smaller firms are less able to substitute across foreign suppliers than large firms, then the former should experience a greater drop in imported quantities than the latter.

To test this hypothesis, I estimate Equations \ref{eq:regression_imports_supplier} and \ref{eq:regression_imports} by partitioning the sample across firms according to their levels of employment. Firms are partitioned according to whether their level of employment is \textit{Above} or \textit{Below} their sector's mean level of employment. As described in Section \ref{sec:data_facts}, I only have a employment and sector level data for a subset of all firms.\footnote{For additional details on the coverage of the employment level data set used in this paper, see Appendix \ref{sec:appendix_features_data}.}
\begin{table}[ht]
    \centering
    \caption{Impact of a Supplier-Supply Shock on Imported Quantities \\ \small By Employment Levels}
    \label{tab:import_instances_empleo_total}
    \footnotesize
\begin{tabular}{lccc | ccc} 
 & \multicolumn{3}{c}{Imported Quantities - $\ln Q_{i,p,d,s,t}$} & \multicolumn{3}{c}{Imported Quantities - $\ln Q_{i,p,d,t}$} \\
 & \textbf{Total} & \textit{Above} & \textit{Below} & \textbf{Total} & \textit{Above} & \textit{Below} \\
 & (1) & (2) & (3) & (4) & (5) & (6)  \\ \hline \hline
 &  &  &  & &  &  \\
$\text{Supplier-Supply-Shock}_{i,t}$ & -0.0713*** & -0.0521* & -0.101***  & -0.0641*** & -0.0692*** & -0.100*** \\
 & (0.0200) & (0.0263) & (0.0273) & (0.0141) & (0.0238) & (0.0276) \\
 &  &  &  &  \\
$\ln Q_{i,p,d}$ lags & YES & YES & YES & YES & YES & YES \\
$\text{Supplier-Supply-Shock}_{i}$ lags & YES & YES & YES & YES & YES & YES \\
$\ln$ RER$_{d,t}$ \& lags & YES & YES & YES & YES & YES & YES \\
$\ln$ Total imported value & YES & YES & YES & YES & YES & YES \\  
 &  &  &  &  \\
Observations & 159,803 & 100,387 & 53,132 & 211,213 & 130,936 & 70,806  \\ \hline \hline
% R-squared & 0.914 & 0.925 & 0.927 & 0.927 \\ \hline
\multicolumn{7}{c}{ Two-way clustered standard errors in parentheses (Firm-Source country level)} \\
\multicolumn{7}{c}{ *** p$<$0.01, ** p$<$0.05, * p$<$0.1} \\
\end{tabular}
\floatfoot{\footnotesize \textbf{Note:} Every specification in the table controls for $\gamma_t$ fixed effects. Columns (1) through (3) also control for $\gamma_{i,p,d,s}$ fixed effects, while columns (4) through (6) control for $\gamma_{i,p,d}$ fixed effects. The specifications which control for lagged values of either the explained and/or the explanatory variables do so by controlling for the $t-1$ and $t-2$ values of the variables. In order to partition firms between \textit{Above} and \textit{Below} I compute the mean level of employment \textit{Above}-\textit{Below} threshold at the ``\textit{year} - \textit{production sector} (at the 6 digit level)'' for firms actively importing.}
\end{table}
Table \ref{tab:import_instances_empleo_total} presents the results for the total sample of firms with employment data and for those firms \textit{above} and \textit{below} their respective sectors' mean employment. Columns (1) through (3) present the results of estimating Equation \ref{eq:regression_imports_supplier}. The estimate on column (1) implies that a one standard deviation supplier-supply shock reduces imported quantities by 1.204\% for the total sample of firms with employment data. Columns (2) and (3) show that firms \textit{below} the mean level of employment at the sector level experience a drop in imported quantities (1.705\% drop) almost twice as a big as firms \textit{above} the this threshold (0.880\%). Columns (4) through (6) repeat the same exercise but by estimating the regression in Equation \ref{eq:regression_imports}. Once more, relatively smaller firms experience a drop in imported quantities almost twice as big as relatively larger firms. Given the results presented in Section \ref{sec:data_facts} which suggest that relatively smaller firms in terms of employment have smaller and/or less diversified network of foreign suppliers, I take these results as evidence that firms with a lower number of linkages with foreign suppliers are more vulnerable to supplier-supply shocks, and experience a greater drop in imported quantities in response to these shocks. Consequently, this result validates the model's prediction that relatively smaller firms exhibit a larger drop in imported quantities than relatively larger firms. 

%%%%%%%%%%%
\subsection{Impact on Exported Quantities} \label{subsec:export_performance}

This section of the paper describes the impact of the supplier-supply shock on firms' exported quantities. To do so, I build on the previous results which show that firms react to an increase in foreign suppliers input costs by reducing imported quantities. Next, I test whether the decrease in imported quantities lead to a reduction in exported quantities.

To test the impact of a supplier-supply shock on firms' exported quantities I estimate the following regression specification
\begin{align} \label{eq:regression_exports}
    \ln QX_{i,p,d,t} = \beta^{X} \text{Supplier-Supply-Shock}_{i,t} + \gamma_{i,p,d} + \gamma_t + \Gamma_{i,p,d,t} + e_{i,p,d,t}
\end{align}
where $\ln QX_{i,p,d,t}$ represents log exported quantities by firm $i$ of product $p$ to destination country $d$ in year $t$, $\gamma_{i,p,d}$ is a firm-product-destination fixed effect which captures invariant features of the export flow of firm $i$ to said foreign market, $\gamma_t$ is a year fixed effect, and $\Gamma_{i,p,d,t}$ represents additional control variables. In particular, I consider the same set of controls as in the regressions carried out in the previous section.

Table \ref{tab:export_instances} presents estimates of parameter $\beta^X$ for several specifications of Equation \ref{eq:regression_exports}. Column (1) shows that not controlling for lagged values of quantities exported or lagged values of the supplier-supply shock leads to an estimated $\beta^{X}$ not significantly different from zero. 
\begin{table}[ht]
    \centering
    \caption{Impact of a Supplier-Supply Shock on Imported Quantities}
    \label{tab:export_instances}
    \small
\begin{tabular}{lcccc} 
 & \multicolumn{4}{c}{Exported Quantities - $\ln QX_{i,p,d,t}$} \\
 & (1) & (2) & (3) & (4) \\ \hline \hline
\\
$\text{Supplier-Supply-Shock}_{i,t}$ & 0.00255 & -0.0345** & -0.0355** & -0.0354** \\
 & (0.0117) & (0.0164) & (0.0164) & (0.0164) \\
 &  &  &  &  \\
$\ln Q_{i,p,d}$ lags & NO & YES & YES & YES \\
$\text{Supplier-Supply-Shock}_{i}$ lags & NO & YES & YES & YES \\
$\ln$ RER$_{d,t}$ \& lags & NO & NO & YES & YES \\
$\ln$ Total imported value & NO & NO & NO & YES \\  
 &  &  &  &  \\
Observations & 415,371 & 153,534 & 137,696 & 137,696 \\ \hline \hline
% R-squared & 0.914 & 0.925 & 0.927 & 0.927 \\ \hline
\multicolumn{5}{c}{ Two-way clustered standard errors in parentheses (Firm-Destination country level)} \\
\multicolumn{5}{c}{ *** p$<$0.01, ** p$<$0.05, * p$<$0.1} \\
\end{tabular}
\floatfoot{\footnotesize \textbf{Note:} Every specification in the table controls for $\gamma_{i,p,d}$ and $\gamma_t$ fixed effects. The specifications which control for lagged values of either the explained and/or the explanatory variables do so by controlling for the $t-1$ and $t-2$ values of the variables. }
\end{table}
However, columns (2) through (4) show that once I control for the lagged values of the supplier-supply shock the impact of a supplier-supply shock is negative and statistically different from zero.\footnote{The lack of statistically significant results in Column (1) may be driven by several causes. For instance, one time periods export instances are frequent in the data. \cite{albornoz2016survival} shows that exporting instances have extremely low survival probabilities. Thus, this result may be capturing firms' learning or experimenting export instances. In Appendix \ref{appendix_regression_results} I show that while controlling for lagged values of quantities exported and lagged values of the supplier-supply shock leads to a significant reduction in the regression observations, these still explain 90\% of total exports. In Appendix \ref{appendix_role_importing} I highlight the role of importing on export performance.} Thus, evidence suggests a positive supplier-supply import price shock leads to a significant reduction in exported quantities. In terms of magnitude, a one standard deviation increase in import prices due to a supplier-supply shock leads to a reduction in exported quantities between 0.582\% and 0.599\%. While the estimated impact is not particularly large, the estimated coefficients are close to those estimated for imported quantities. In other words, these results suggests provide both further evidence of the imperfect substitutability between foreign suppliers, and evidence of a significant role of imported inputs in export performance, validating the model's predictions. 

Next, I repeat the estimation of Equation \ref{eq:regression_exports} 
but partitioning the sample depending on firms' employment. As in Section \ref{subsec:import_performance}, this partition is carried out to emphasize the role of imported inputs in exporting and, more precisely, the significance of granular foreign supplier linkages.
\begin{table}[ht]
    \centering
    \caption{Impact of a Supplier-Supply Shock on Exported Quantities \\ \footnotesize By Employment Level}
    \label{tab:export_instances_empleo_total}
    \small
\begin{tabular}{l c c c} 
 & \multicolumn{3}{c}{Exported Quantities - $\ln QX_{i,p,d,t}$} \\
 & \textbf{Total} & \textit{Above} & \textit{Below} \\
 & (1) & (2) & (3) \\ \hline \hline
\\
$\text{Supplier-Supply-Shock}_{i,t}$ & -0.0342* & -0.00454 & -0.0751* \\
 & (0.0177) & (0.0297) & (0.0387) \\
 &  &  &  \\
$\ln Q_{i,p,d}$ lags & YES & YES & YES \\
$\text{Supplier-Supply-Shock}_{i}$ lags & YES & YES & YES \\
$\ln$ RER$_{d,t}$ \& lags & YES & YES & YES \\
$\ln$ Total imported value & YES & YES & YES \\  
 &  &  &  \\
Observations & 118,397 & 75,421 & 38,175 \\ \hline \hline
% R-squared & 0.914 & 0.925 & 0.927 & 0.927 \\ \hline
\multicolumn{4}{c}{ Two-way clustered standard errors in parentheses (Firm-Destination country level)} \\
\multicolumn{4}{c}{ *** p$<$0.01, ** p$<$0.05, * p$<$0.1} \\
\end{tabular}
\floatfoot{\footnotesize \textbf{Note:} Every specification in the table controls for $\gamma_{i,p,d}$ and $\gamma_t$ fixed effects. The specifications which control for lagged values of either the explained and/or the explanatory variables do so by controlling for the $t-1$ and $t-2$ values of the variables. In order to partition firms between \textit{Above} and \textit{Below} I compute the mean level of employment \textit{Above}-\textit{Below} threshold at the ``\textit{year} - \textit{production sector} (at the 6 digit level)'' for firms actively exporting. }
\end{table}
Table \ref{tab:export_instances_empleo_total} presents the results of this empirical exercise. Column (1) shows that the estimated parameter $\beta^X$ for a sample of firms with employment information is approximately to the result presented in Table \ref{tab:export_instances} (-0.0342 vs. -0.0354). On the one hand, column (2) shows that firms with levels of employment \textit{above} the sector's mean experience no significant drop in exported quantities in response to a supplier-supply shock. On the other hand, column (3) shows that firms with levels of employment \textit{below} the sector's mean experience a drop in exported quantities twice as big as the in the full sample. This result is line with the results found in Table \ref{tab:import_instances_empleo_total} which show that relatively smaller firms also experience a greater drop in imported quantities than relatively larger firms. Comparing the estimated coefficient in column (3) of Table \ref{tab:export_instances_empleo_total} with the coefficients presented in columns (3) and (6) of Table \ref{tab:import_instances_empleo_total}, a supplier-supply shock that reduces imported quantities by 10\% lead to a reduction in exported quantities of approximately 7.5\%. Once again, these results validate the predictions of the model presented in Section \ref{sec:model}, by suggesting both a strong imported input to export complementarities, and relatively smaller firms being hit harder by a supplier-supply shock.

\noindent
\textbf{Robustness check.} The shocks constructed in Section \ref{subsec:supplier_supply_shocks} compute the change in import prices as the log differences of prices. An alternative measure of changes in prices is the percentage difference, or the difference between the prices in periods $t$ and $t+1$ divided by their average
\begin{align} \label{eq:change_prices_percentage_differences}
    \Delta  \tilde{P}_{i,k,t} = \frac{ P_{i,k,t} - P_{i,k,t-1}}{0.5 \times \left(P_{i,k,t} + P_{i,k,t-1} \right)}
\end{align}
I re-estimate equation \ref{eq:price_reg_supplier} as
\begin{align} \label{eq:price_reg_supplier_percentage_differences}
    \Delta \tilde{P}_{i,p,d,s,t} = \tilde{\beta}_{i,t} + \tilde{\gamma}_{s,t} + \tilde{\epsilon}_{i,p,d,s,t}
\end{align}
and construct the new Supplier-Supply-Shocks
\begin{align} \label{eq:shock_second_specification}
    \tilde{\text{Supplier-Supply-Shock}}_{i,t} = \sum_s \omega_{s,t-1} \times \tilde{\gamma}_{s,t}
\end{align}

In Appendix \ref{appendix_regression_results}, I show that under this shock specification Supplier-Supply Shocks constructed using changes in prices as the percentage difference lead to a significant reduction in imported quantities.\footnote{See Tables \ref{tab:import_instances_supplier_pdiff} through \ref{tab:import_instances_pdiff_empleo_total} in Appendix \ref{appendix_regression_results}.} Next, I show that the impact of supplier-supply shocks in exported quantities is robust to this shock specification. 
\begin{table}[ht]
    \centering
    \caption{Impact of Supplier-Supply Shock on Exported Quantities \\ \footnotesize Robustness Check}
    \label{tab:export_instances_pdiff}
    \small
\begin{tabular}{lcccc} 
 & \multicolumn{4}{c}{Exported Quantities - $\ln QX_{i,p,d,t}$} \\
 & (1) & (2) & (3) & (4) \\ \hline \hline
\\
$\tilde{\text{Supplier-Supply-Shock}}_{i,t}$ & -0.00913 & -0.145*** & -0.134*** & -0.136*** \\
 & (0.0279) & (0.0497) & (0.0502) & (0.0500) \\
 &  &  &  &  \\
$\ln Q_{i,p,d}$ lags & NO & YES & YES & YES \\
$\tilde{\text{Supplier-Supply-Shock}}_{i}$ lags & NO & YES & YES & YES \\
$\ln$ RER$_{d,t}$ \& lags & NO & NO & YES & YES \\
$\ln$ Total imported value & NO & NO & NO & YES \\  
 &  &  &  &  \\
Observations & 415,371 & 153,534 & 137,696 & 137,696 \\ \hline \hline
% R-squared & 0.914 & 0.925 & 0.927 & 0.927 \\ \hline
\multicolumn{5}{c}{ Two-way clustered standard errors in parentheses (Firm-Destination country level)} \\
\multicolumn{5}{c}{ *** p$<$0.01, ** p$<$0.05, * p$<$0.1} \\
\end{tabular}
\floatfoot{\footnotesize \textbf{Note:} Every specification in the table controls for $\gamma_{i,p,d}$ and $\gamma_t$ fixed effects. The specifications which control for lagged values of either the explained and/or the explanatory variables do so by controlling for the $t-1$ and $t-2$ values of the variables. }
\end{table}
Table \ref{tab:export_instances_pdiff} presents estimates of semi-elasticity $\beta^X$ from Equation \ref{eq:regression_exports} using the supplier-supply-shock specified by Equation \ref{eq:shock_second_specification}. In line with the results under the first shock specification in Table \ref{tab:export_instances}, a positive supplier-supply shock which increases firms' cost of imported bundles leads to a significant reduction in exported quantities. In terms of magnitude, a one standard deviation of shock $\tilde{\text{Supplier-Supply-Shock}}_{i,t}$ of 0.0847 leads to a reduction of exported quantities between 1.15\% and 1.22\%.\footnote{For details and descriptive statistics of the constructed shocks see Appendix \ref{sec:appendix_shocks_desc}.} The estimated impact is approximately twice as large as the one estimated for the first shock specification, presented in Table \ref{tab:export_instances}.

Finally, I test whether the impact of a supplier-supply shock is greater for relatively smaller firms under the alternative shock specification. Table \ref{tab:export_instances_pdiff_empleo_total} presents estimated values of semi-elasticity $\beta^X$ for different sample partitions across firms' size.
\begin{table}[ht]
    \centering
    \caption{Impact of Supplier-Supply Shock on Exported Quantities \\ \footnotesize By Employment - Robust.}
    \label{tab:export_instances_pdiff_empleo_total}
    \small
\begin{tabular}{l c c c} 
 & \multicolumn{3}{c}{Exported Quantities - $\ln QX_{i,p,d,t}$} \\
 & \textbf{Total} & \textit{Above} & \textit{Below} \\
 & (1) & (2) & (3) \\ \hline \hline
\\
$\tilde{\text{Supplier-Supply-Shock}}_{i,t}$ & -0.149*** & -0.0789 & -0.215*** \\
 & (0.0482) & (0.0644) & (0.0783) \\
 &  &  &  \\
$\ln Q_{i,p,d}$ lags & YES & YES & YES \\
$\text{Supplier-Supply-Shock}_{i}$ lags & YES & YES & YES \\
$\ln$ RER$_{d,t}$ \& lags & YES & YES & YES \\
$\ln$ Total imported value & YES & YES & YES \\  
 &  &  &  \\
Observations & 118,397 & 75,421 & 38,175 \\ \hline \hline
% R-squared & 0.914 & 0.925 & 0.927 & 0.927 \\ \hline
\multicolumn{4}{c}{ Two-way clustered standard errors in parentheses (Firm-Destination country level)} \\
\multicolumn{4}{c}{ *** p$<$0.01, ** p$<$0.05, * p$<$0.1} \\
\end{tabular}
\floatfoot{\footnotesize \textbf{Note:} Every specification in the table controls for $\gamma_{i,p,d}$ and $\gamma_t$ fixed effects. The specifications which control for lagged values of either the explained and/or the explanatory variables do so by controlling for the $t-1$ and $t-2$ values of the variables. In order to partition firms between \textit{Above} and \textit{Below} I compute the mean level of employment \textit{Above}-\textit{Below} threshold at the ``\textit{year} - \textit{production sector} (at the 6 digit level)'' for firms actively exporting. }
\end{table}
Column (1) shows that the estimated coefficient for the sample of firms for which I have employment information is close to that presented in Table \ref{tab:export_instances_pdiff} (-0.149 vs -0.136, respectively). On the one hand, column (2) shows that a supplier-supply shock does not have a significant impact on export performance of relatively larger firms, i.e., firms above the mean level of sector employment. On the other hand, column (3) shows that relatively smaller firms' exported quantities are significantly and negatively impacted by a positive supplier-supply shock. The estimated impact is approximately 50\% larger than for the full sample estimate.

In summary, this section of the paper constructed supplier-supply shock by exploiting the additional dimension of information provided by the data set's foreign supplier information. In particular, I identify supplier-time specific changes in domestic firms' price of imported inputs through a battery of fixed effects. In order to construct firm-level supplier-supply shocks I weigh the recovered supplier-specific changes in import prices by firm network's lagged exposure to foreign suppliers. The constructed shocks are associated with a significant drop in both imported and exported quantities. In line with the stylized facts presented in Section \ref{subsec:stylized_facts}, relatively smaller firms which have smaller and/or less diversified networks of foreign suppliers are particularly hit by supplier-supply shocks. Furthermore, I show that these results are robust to a different shock specification. Overall, the presented suggest both the presence of search frictions in the market for foreign suppliers and the presence of significant imported input to export performance complementarities, in line with the predictions of the theoretical framework introduced in Section \ref{sec:model}.
    
%%%%%%%%%%%%%%%%%%%%%%%%%%%%%%%%%%%%%%%%%%%%%%%%%%%%%%%%%%%%%%%%%
\section{Conclusion} \label{sec:conclusion}

This paper uses detailed customs level data to present novel evidence on the granular nature of domestic firms' network of foreign suppliers. I show that the distribution of the number of firms' foreign suppliers is highly skewed, with the median domestic firm only buying from two foreign suppliers. While relatively larger and  exporting firms have a greater number of foreign suppliers, I show that even these types of firms tend to focus a significant share of their imported value on only one foreign supplier. I complement these stylized facts with evidence of significant search and informational frictions in international markets for foreign suppliers.

In order to guide my empirical work I construct a theoretical framework of heterogeneous firms subject to search and matching frictions in the market for foreign suppliers. I parametrize and calibrate the model to match the novel stylized facts and other moments from the data and show that the framework yields two sharp predictions: (i) relatively smaller or less productive firms experience larger drops in imported and exported quantities in response to an increase in the cost of imported inputs triggered by a supplier shock than relatively larger or more productive firms, (ii) strong complementarities between imported inputs and export performance.

I take the theoretical framework to the data by constructing firm specific supplier supply cost shocks using a shift-share analysis. I identify the variation in import prices due to supplier specific components through a battery of fixed effects and aggregating them using firms' lagged exposure to foreign suppliers. Results show that an increase in the cost of imported inputs explained by a supplier-supply shock leads to a significant drop in imported and exported quantities, particularly for relatively small firms. Additionally, these results suggest strong imported-input to export complementarities, validating the predictions of my theoretical framework. 

\textbf{Acknowledgements:} \normalsize

I would also like to thank the comments of Matthias Doepke, Guido Lorenzoni and Kiminori Matsuyama. I would like to thank the attendees of Northwestern University's macro lunch seminar, Fergal Hanks and Yong Cai. Susan Belles provided helpful comments.

\newpage
\bibliography{biblio.bib}

\newpage
\appendix

\section{Appendix Data Cleaning} \label{sec:Appendix}

\subsection{Cleaning procedure of supplier identification} \label{subsec:Appendix_cleaning}

\noindent In this appendix, I explain the algorithm to clean suppliers' name variable. This is the same algorithm used in \cite{berninietal2021}. Most of the problems detected were typing errors or different ways to write the same firm. For example, "Volkswagen" could be find as (a) "Volkswagen", (b) "Volkswagen international", (c) "VW", (d) "Volkswgen", etc. If I ignore this problem, I could interpret them as 4 different suppliers, when they actually are the same company.

\noindent In order to fix this problem, I apply the following procedure:

\begin{itemize}
    \item $1^{st}$ step: I delete special characters (e.g. "/" and "-"), double spaces, country names (e.g. United States) and their acronym (e.g. USA), and companies' suffixes (LLC, LTD, SA).
    \item $2^{nd}$ step: If a suppliers' name includes the name of a famous multinational firm (or its acronym), I replace the supplier name with them. For example, if a supplier includes Volkswagen or VW, I define them as Volkswagen.
    \item $3^{rd}$ step: I compare the similarity of importers and suppliers names using \cite{raffo2009play} bigram technique (I explain in detail below). If this technique report high similarity, I replace suppliers' name by the importer's name.
    \item $4^{th}$ step: I applied the same procedure of step 3 to compare all suppliers within each firm and determine which are the same company.
\end{itemize}

\noindent The $1^{st}$ step of the procedure allows as to fix cases as previous (b) example by only deleting common words that are not the core name of the company. $2^{nd}$ step is design to solve (c) problems, where a commonly known firm could be rightly write by different ways. 

\noindent However, typing errors, as the example (d), are more difficult to detect and fix. I use \cite{raffo2009play} bigram technique, which compare two string variables and split them by a sequence of two characters called "bigram set" (e.g. "Ford" bigram set is [Fo, or, rd]). The algorithm compute a similscore as the ratio between the union of both bigram sets and all components of both sets. In addition, I use two different weights to account to common bigrams: Simple and log weights. 

\noindent Hence, in $3^{rd}$ and $4^{th}$ step I obtain two similscores in each step, then I define that two strings has "high similarity" if both scores are above 0.65, or one is above 0.8 and the other above 0.35.

%%%%%%%%%%%%%%%%%%%%%%%%%%%%%%%%%%%%%%%%%%%%%%%%%%%%%%%%%%%%%
\newpage
\section{Additional Information Stylized Facts} \label{sec:appendix_stylized_facts}

In this section of the paper I present additional information on the stylized facts presented in Section \ref{subsec:stylized_facts}. First, I start by describing the distribution of domestic firms' number of foreign suppliers shown in Figure \ref{fig:Histogram_Number_Suppliers}. Table \ref{tab:appendix_dist_number_suppliers} presents statistics of this distribution.
\begin{table}[ht]
    \centering
    \caption{Distribution of Domestic Firms' Number of Foreign Suppliers \\ \footnotesize Descriptive Statistics}
    \label{tab:appendix_dist_number_suppliers}
    \footnotesize 
    \begin{tabular}{l c c c c c c}
                &  Mean & Median & p75 & p90 & p95 & p99 \\ \hline \hline \\
Total Sample    & 6.11 & 2 & 5 & 13 & 23 & 64 \\
\\
Not-Exporters   & 3.83 & 2 & 4 & 8 & 13& 31 \\
Exporters       & 11.33 & 4 & 10 & 27 & 46 & 122 \\
\textit{Above} Mean Exported Value & 36.97 & 16 & 46 & 91 & 143.5 & 301 \\
\\
Firms with Employment Data & 16.90 & 7 & 17 & 39 & 62 & 160 \\
\textit{Below} Mean Employment & 9.37 & 5 & 11 & 21 & 31 & 62 \\
\textit{Above} Mean Employment & 32.46 & 16 & 37 & 75 & 112 & 257 \\
\\ \hline \hline
    \end{tabular}
\end{table}

Next, Table \ref{tab:appendix_dist_share_top} presents statistics on the distribution of the share of domestic firms' imported value by their top-supplier. 
\begin{table}[ht]
    \centering
    \caption{Distribution of Share of Firms' Total Value Imported \\ \footnotesize Explained by Top Supplier}
    \label{tab:appendix_dist_share_top}
    \footnotesize
    \begin{tabular}{l c c c c c c}
        &  Mean & p10 & p25 & p50 & p75 & p90 \\ \hline \hline \\
Total Sample &  .746 & .335 & .520 & .832 & 1 & 1  \\
\\
Not-Exporters & .781 & .381 & .567 & .922 & 1 & 1  \\
Exporters     & .666 & .270 & .420 & .667 & 1 & 1   \\
\textit{Above} Mean Exported Value & .769 & .359 & .549 & .893 & 1 & 1 \\
\\
Firms with Employment Data & .732 & .323 & .505 & .799 & 1 & 1 \\
\textit{Below} Mean Employment & .745 & .340 & .520 & .825 & 1 & 1 \\
\textit{Above} Mean Employment & .493 & .192 & .285 & .443 & .677 & .901 \\ \\ \hline \hline
    \end{tabular}
\end{table}
In line with the results presented in Section \ref{subsec:stylized_facts}, the distribution is highly skewed towards higher values.  

\noindent
\textbf{Additional details on search and matching stylized facts.} Next, I turn to providing additional details on the presence of search and matching frictions in the market for foreign suppliers.

I start analyzing the dynamic properties of international firm-to-firm linkages by describing their survival probability. Table \ref{tab:SF_survival} presents a regression analysis on the probability of a linkage being active in years $t$ and $t+1$. 
\begin{table}[ht]
    \centering
    \caption{Unconditional and Conditional Survival Probability \\ Top Supplier}
    \footnotesize
    \label{tab:SF_survival}
\begin{tabular}{lcccccc}
 & \multicolumn{6}{c}{Probability of Linkage being active in $t$ and $t+1$} \\ 
 & (1) & (2) & (3) & (4) & (5) & (6) \\ \hline \hline
 &  &  &  &  &  &  \\
Constant & 0.404*** & 0.513*** & 0.396*** & 0.409*** & 0.562*** & 0.366*** \\
 & (0.000463) & (0.00117) & (0.000452) & (0.000447) & (0.00101) & (0.000496) \\
Top Supplier &  &  & 0.130*** &  &  & 0.277*** \\
 &  &  & (0.00125) &  &  & (0.00150) \\
 &  &  &  &  &  &  \\
Top Supplier & NO & YES & NO & NO & YES & NO \\
Firm FE      & NO & NO  & NO & YES & YES & YES \\
Year FE      & NO & NO  & NO & YES & YES & YES \\
&  &  &  &  &  &  \\
Observations & 1,121,213 & 183,319 & 1,121,213 & 1,107,124 & 165,591 & 1,107,124 \\ \hline \hline
% R-squared & 0.000 & -0.000 & 0.010 & 0.116 & 0.457 & 0.143 \\ \hline \hline
% R-squared & -0.000 & 0.000 & 0.010 & 0.112 & 0.442 & 0.139 \\ \hline
\multicolumn{7}{c}{ Standard errors in parentheses} \\
\multicolumn{7}{c}{ *** p$<$0.01, ** p$<$0.05, * p$<$0.1} \\
\end{tabular}
\end{table}
Column (1) shows that the unconditional survival probability is 0.404. Column (2) runs the same empirical analysis but conditions the sample to only include firms' top supplier linkages. The survival probability of this type of linkages is 0.513. Column (3) runs the same analysis for the whole sample but introduces a binary control variable which takes the value of one for a firms' top supplier linkage. The coefficient associated with this binary control variable implies that a top supplier's survival probability is 13 percentage points greater. Columns (4) through (6) run the same specification but controlling for firm and year fixed effects. While the unconditional probability varies little when controlling for these fixed effects, the additional survival probability of a top supplier linkage is higher when introducing these fixed effect (13 versus 27.7 percentage points).

Next, I explore whether linkages between a domestic firm and its top supplier is persistence across time. In order to test this persistence I compute the probability of a firm's top foreign supplier in year $t$ remaining as top supplier in year $t+j$ for $j=\{1,2,3 \}$. Table \ref{tab:SF_persistence} presents the results of a regression analysis computing these probabilities conditioning on the linkage still being active.
\begin{table}[ht]
    \centering
    \caption{Persistence of Top Supplier Linkage}
    \label{tab:SF_persistence}
        \footnotesize
\begin{tabular}{lcccccc} 
 & \multicolumn{2}{c}{Top Supplier $t$ \& $t+1$} & \multicolumn{2}{c}{Top Supplier $t$ \& $t+2$} & \multicolumn{2}{c}{Top Supplier $t$ \& $t+3$} \\ 
 & (1) & (2) & (3) & (4) & (5) & (6) \\ \hline \hline
 &  &  &  &  &  &  \\
Constant & 0.716*** & 0.710*** & 0.693*** & 0.693*** & 0.679*** & 0.683*** \\
 & (0.00147) & (0.00140) & (0.00182) & (0.00163) & (0.00208) & (0.00178) \\
 &  &  &  &  &  &  \\
Firm FE      & NO & YES & NO & YES & NO & YES \\
Year FE      & NO & YES & NO & YES & NO & YES \\
 &  &  &  &  &  &  \\
Observations & 93,985 & 84,982 & 63,995 & 59,270 & 50,519 & 47,070 \\ \hline \hline
% R-squared & 0.000 & 0.338 & -0.000 & 0.405 & 0.000 & 0.464 \\ \hline
\multicolumn{7}{c}{ Standard errors in parentheses} \\
\multicolumn{7}{c}{ *** p$<$0.01, ** p$<$0.05, * p$<$0.1} \\
\end{tabular}
\end{table}
Columns (1) and (2) show the unconditional and conditional on firm and year fixed effects probabilities for year $t+1$ which are approximately 70\%. Columns (3) through (6) show that the probability of a domestic firm having the same top foreign supplier two and three years into the future remain close to 70\%.

Finally, I turn to analyzing the success probability of the search for new suppliers. Success in this search is defined as the probability of the linkage to remain active in the year after the initial interaction between firms. Table \ref{tab:SF_survival1_entrant} carries out an analogous regression analysis to Table \ref{tab:SF_survival}, focusing on new suppliers. It is straightforward that the success probability of a new supplier is significantly lower than the unconditional survival probability.
\begin{table}[ht]
    \centering
    \caption{Unconditional and Conditional Linkage Probability \\ Entrants}
    \label{tab:SF_survival1_entrant}
    \footnotesize
\begin{tabular}{lcccccc}
 & \multicolumn{6}{c}{Probability of Linkage being active in $t$ and $t+1$} \\ 
 & (1) & (2) & (3) & (4) & (5) & (6) \\ \hline \hline
 &  &  &  &  &  &  \\
Constant & 0.404*** & 0.274*** & 0.487*** & 0.409*** & 0.276*** & 0.515*** \\
 & (0.000463) & (0.000675) & (0.000579) & (0.000447) & (0.000667) & (0.000594) \\
Entrant &  &  & -0.213*** &  &  & -0.275*** \\
 &  &  & (0.000929) &  &  & (0.00105) \\    
 &  &  &  &  &  &  \\
Entrant & NO & YES & NO & NO & YES & NO \\
Firm FE      & NO & NO  & NO & YES & YES & YES \\
Year FE      & NO & NO  & NO & YES & YES & YES \\
&  &  &  &  &  &  \\
Observations & 1,121,213 & 436,146 & 1,121,213 & 1,107,124 & 422,958 & 1,107,124 \\ \hline \hline
% R-squared & -0.000 & 0.000 & 0.055 & 0.112 & 0.108 & 0.170 \\ \hline
\multicolumn{7}{c}{ Standard errors in parentheses} \\
\multicolumn{7}{c}{ *** p$<$0.01, ** p$<$0.05, * p$<$0.1} \\
\end{tabular}
\end{table}
Column (2) computes the survival probability only on a sample of new suppliers, exhibiting an estimate 0.276 compared 0.418 in Column (1). Column (3) runs the same regression by controlling for the dummy variable "Entrant" and finds a significant negative coefficient. Columns (4) through (6) carry out the same regressions but add firm and year fixed effects. Across specifications and sample partitions, the success probability of new suppliers is close to being half of the unconditional estimate.  

%%%%%%%%%%%%%%%%%%%%%%%%%%%%%%%%%%%%%%%%%%%%%%%%%%%%%%%%%%%%%
\newpage
\section{Additional Features of the Data} \label{sec:appendix_features_data}

In this section of the appendix I present additional of the data, in particular, I present statistics which describe the distribution of domestic firms' exports and imports per firm, and the distribution of foreign suppliers exports per firm
First, I present Tables \ref{tab:dist_imports_per_firm} to \ref{tab:dist_exports_per_firm_domestic} which describe the distribution of value imported per domestic firm, value exported per foreign supplier firm and value exported per domestic firm, respectively.
\begin{table}[ht]
    \centering
    \caption{Distribution of Value Exported by Domestic Firm}
    \label{tab:dist_exports_per_firm_domestic}
    \tiny
    \begin{tabular}{c|c|c|c|c|c|c|c|c|c}
              & Mean & p5 & p10 & p25 & p50 & p75 & p90 & p95 & p99  \\ \hline \hline
         2000 & 2,300,131 & 1,168 & 2,932 & 9,620 & 45,986 & 247,517 & 1,374,957 & 3,889,610 & 32,998,131 \\
         2001 & 2,390,581 & 1,250 & 2,742 & 9,542 & 44,222 & 248,067 & 1,342,529 & 3,979,841 & 33051508 \\
         2002 & 1,982,188 & 730 & 1,674 & 6,000 & 27,912 & 160,948 & 971,451 & 2,871,974 & 23,870,399 \\
         2003 & 2,435,238 & 1,025 & 2,287 & 7,692 & 34,168 & 193,414 & 1,168,295 & 3,543,958 & 26,534,729 \\
         2004 & 2,435,238 & 1,025 & 2,287 & 7,692 & 34,168 & 193,414 & 1,168,295 & 3,543,958 & 26,534,729 \\
         2005 & 2,680,331 & 1,250 & 2,639 & 8,800 & 39,404 & 228,693 & 1,350,542 & 3,930,673 & 28,613,176 \\
         2006 & 3,068,924 & 1,469 & 3,078 & 10,086 & 46,898 & 265,201 & 1,618,939 & 4,710,610 & 33,390,779 \\
         2007 & 3,832,875 & 2,024 & 4,120 & 13,273 & 61,853 & 370,335 & 2,138,773 & 5,991,423 & 36,793,402 \\
         2008 & 4,871,963 & 2,450 & 4,899 & 16,486 & 74,809 & 466,926 & 2,571,928 & 7,478,486 & 44,167,865 \\
         %2009 & 4,118,082 & 2,500 & 4,966 & 16,000 & 72,401 & 438,349 & 2,462,797 & 7,105,613 & 42,639,837 \\
         %2010 & 5,158,410 & 3,034 & 5,640 & 18,519 & 84,183 & 507,263 & 2,852,070 & 7,774,306 & 51,329,965 \\
%         2011 & 6,389,458 & 3,250 & 6,496 & 22,593 & 100,371 & 632,771 & 3,509,297 & 9,681,199 & 62,542,782 \\
%         2012 & 6,337,445 & 3,572 & 7,188 & 24,801 & 110,540 & 667,862 & 3,580,635 & 9,886,407 & 66,553,233 \\
%         2013 & 6,552,489 & 3,760 & 7,543 & 26,153 & 119,894 & 728,456 & 3,674,421 & 10,097,067 & 73,868,836 \\
%         2014 & 6,437,256 & 3,500 & 6,847 & 23,406 & 107,388 & 677,723 & 3,554,678 & 10,000,000 & 80,240,950 \\
%         2015 & 5,947,388 & 3,500 & 6,921 & 21,749 & 101,969 & 644,289 & 3,472,075 & 9,796,250 & 78,005,307 \\
%         2016 & 5,984,360 & 2,960 & 5,950 & 20,001 & 92,267 & 612,965 & 3,514,213 & 9,302,190 & 80,890,961 \\
    \end{tabular}
\end{table}
\begin{table}[ht]
    \centering
    \caption{Distribution of Value Imported per Firm}
    \label{tab:dist_imports_per_firm}
    \tiny
    \begin{tabular}{c|c|c|c|c|c|c|c|c|c}
              & Mean & p5 & p10 & p25 & p50 & p75 & p90 & p95 & p99  \\ \hline \hline
         2000 & 920,500 & 910 & 2,670 & 11,230 & 49,125 & 236,965 & 1,027,424 & 2,445,182 & 14,169,365 \\
         2001 & 810,349 & 895 & 2,489 & 10,178 & 45,188 & 213,848 & 900,184 & 2,264,536 & 13,974,765 \\
         2002 & 555,383 & 450 & 1,357 & 6,012 & 26,540 & 125,914 & 567,231 & 1,537,635 & 9,781,031 \\
         2003 & 719,020 & 621 & 1,800 & 7,755 & 36,885 & 180,887 & 810,211 & 2,041,917 & 11,737,183 \\
         2004 & 1,006,572 & 625 & 1,958 & 9,060 & 43,861 & 221,329 & 969,482 & 2,439,342 & 15,402,843 \\
         2005 & 1,183,486 & 500 & 1,575 & 8,234 & 45,521 & 236,030 & 1,050,825 & 2,679,588 & 17,510,612 \\
         2006 & 1,369,051 & 438 & 1,633 & 9,525 & 49,856 & 260,259 & 1,190,456 & 2,984,414 & 19,501,204 \\
         2007 & 1,843,175 & 890 & 2,836 & 14,490 & 71,503 & 356,030 & 1,619,753 & 3,943,917 & 26,091,518 \\
         2008 & 2,318,793 & 1,259 & 3,613 & 17,910 & 85,603 & 420,694 & 1,863,988 & 4,556,231 & 29,668,393 \\
%         2009 & 1,677,279 & 1,311 & 3,639 & 16,508 & 78,269 & 373,710 & 1,596,218 & 3,808,024 & 23,856,241 \\
%         2010 & 2,427,821 & 1,536 & 4,299 & 19,599 & 95,390 & 465,623 & 2,076,565 & 5,047,004 & 32,219,735 \\
%         2011 & 3,135,206 & 1,907 & 5,125 & 25,272 & 116,750 & 572,584 & 2,453,422 & 6,045,922 & 39,307,590 \\
    \end{tabular}
\end{table}
\begin{table}[ht]
    \centering
    \caption{Distribution of Value Exported by Foreign Firm}
    \label{tab:dist_exports_per_firm_foreign}
    \tiny
    \begin{tabular}{c|c|c|c|c|c|c|c|c|c}
              & Mean & p5 & p10 & p25 & p50 & p75 & p90 & p95 & p99  \\ \hline \hline
         2000 & 231,785 & 292 & 1,100 & 5,300 & 19,689 & 67,654 & 232,491 & 530,402 & 2,842,062 \\
         2001 & 202,523 & 318 & 1,151 & 5,095 & 18,250 & 60,837 & 216,906 & 489,716 & 2,635,970 \\
         2002 & 152,034 & 143 & 680 & 3,509 & 13,358 & 45,762 & 160,220 & 369,282 & 2,194,979 \\
         2003 & 185,239 & 200 & 888 & 4,074 & 14,618 & 51,391 & 191,357 & 438,999 & 2,435,167 \\
         2004 & 245,992 & 200 & 963 & 4,644 & 17,198 & 59,903 & 217,542 & 511,912 & 3,003,776 \\
         2005 & 276,593 & 64 & 597 & 3,719 & 16,428 & 60,124 & 224,321 & 540,181 & 3,186,962 \\
         2006 & 305,874 & 39 & 382 & 3,600 & 16,712 & 63,077 & 234,545 & 567,377 & 3,436,139 \\
         2007 & 378,955 & 50 & 515 & 4,309 & 19,876 & 74,718 & 280,399 & 672,429 & 4,142,583 \\
         2008 & 469,384 & 90 & 828 & 5,107 & 23,293 & 87,830 & 324,319 & 778,303 & 4,976,328 \\
%         2009 & 348,047 & 80 & 750 & 4,715 & 20,872 & 76,001 & 273,667 & 639,974 & 3,926,958 \\
%         2010 & 481,421 & 150 & 1,104 & 5,716 & 24,535 & 90,365 & 338,319 & 806,740 & 4,839,376 \\
%         2011 & 615,984 & 203 & 1,277 & 6,687 & 29,260 & 109,214 & 414,704 & 972,718 & 5,793,553 \\
    \end{tabular}
\end{table}
Tables \ref{tab:dist_exports_per_firm_domestic} and \ref{tab:dist_imports_per_firm} show that the distribution of exports per firm is highly more concentrated than the distribution of imports per firm. This is primarily explained by the significant small share of exporting firms. Interestingly, the distribution of value exported by foreign supplier to Argentina is relatively less concentrated. There are several reasons which may explain this result. First, the set of global firms which can sell to Argentinean firms is incommensurate, providing domestic firms with a wide portfolio of potential suppliers. Second, the left tail of the distribution of exports of foreign suppliers per firm is in line with the high-rate of search and high-supplier turnover presented in Tables \ref{fig:Search_Evidence}. Thus, low exports per foreign supplier can be explained by domestic firms search and experimenting the quality of a new linkage. 

\noindent
\textbf{Additional Details on Foreign Suppliers Performance.} Next, I present additional descriptive statistics on foreign suppliers' export performance to Argentinean firms. To do so, I describe foreign suppliers' distribution of export instances defined at the ``domestic Argentinean firm''-``product HS-6''-``source country'' level, at the ``product HS-6'' level, at the ``source country'' level, and at the ``domestic Argentinean firm'' level (or customer level). Furthermore, I show the share of total value exported by foreign suppliers to Argentinean firms (or total value imported by Argentinean firms) for foreign suppliers in different quantiles of these distributions.
\begin{table}[ht]
    \centering
    \caption{Foreign Suppliers' Distribution of Export Instances}
    \label{tab:fs_dist_instances}
    \tiny
    \begin{tabular}{c|c|c|c|c|c|c|c}
    \multicolumn{8}{c}{ } \\
    \multicolumn{8}{c}{Number of Export Instances per Foreign Supplier} \\ \hline \hline
	&	Mean	&	Median	&	p75	&	p90	&	p95	&	p99	&	Max	\\ \hline \hline
2000	&	6.16	&	2	&	5	&	12	&	21	&	65	&	 2,314 	\\
2001	&	5.80	&	2	&	5	&	11	&	20	&	61	&	 2,028 	\\
2002	&	5.31	&	2	&	4	&	10	&	18	&	59	&	 1,390 	\\
2003	&	5.57	&	2	&	4	&	11	&	19	&	62	&	 1,982 	\\
2004	&	5.75	&	2	&	4	&	11	&	20	&	63	&	 2,340 	\\
2005	&	5.78	&	2	&	4	&	11	&	20	&	63	&	 2,776 	\\
2006	&	5.76	&	2	&	4	&	11	&	20	&	62	&	 2,621 	\\
2007	&	5.85	&	2	&	4	&	11	&	20	&	64	&	 3,432 	\\
2008	&	5.74	&	2	&	4	&	11	&	20	&	62	&	 3,534 	\\ \hline \hline
    \end{tabular} \\
    \centering
    \tiny
    \begin{tabular}{c|c|c|c|c|c}
    \multicolumn{6}{c}{ } \\
    \multicolumn{6}{c}{Share of Exported Value by Foreign Suppliers by Distribution of \# Instances} \\ \hline \hline
	&	Bottom 40\%	&	40\% to 60\%	&	60\% to 80\%	&	Top 20\%	&	Top 5\%	\\ \hline \hline
2000	&	8.53	&	9.93	&	9.25	&	72.28	&	51.08	\\
2001	&	9.41	&	11.15	&	9.45	&	69.99	&	50.49	\\
2002	&	12.23	&	8.10	&	11.62	&	68.05	&	48.25	\\
2003	&	10.12	&	6.80	&	12.39	&	70.69	&	49.73	\\
2004	&	9.46	&	7.12	&	13.99	&	69.42	&	52.63	\\
2005	&	8.21	&	6.79	&	14.69	&	70.31	&	52.45	\\
2006	&	8.20	&	6.44	&	13.02	&	72.34	&	54.94	\\
2007	&	8.61	&	5.91	&	13.34	&	72.14	&	54.58	\\
2008	&	8.83	&	7.38	&	13.27	&	70.53	&	53.52	\\ \hline \hline
    \end{tabular}
\end{table}
\begin{table}[ht]
    \centering
    \caption{Foreign Suppliers' Distribution of Exported Products}
    \label{tab:fs_dist_products}
    \tiny
    \begin{tabular}{c|c|c|c|c|c|c|c}
    \multicolumn{8}{c}{ } \\
    \multicolumn{8}{c}{Number of Exported Products per Foreign Supplier} \\ \hline \hline
	&	Mean	&	Median	&	p75	&	p90	&	p95	&	p99	&	Max	\\ \hline \hline
2000	&	 4.78 	&	2	&	4	&	10	&	17	&	49	&	372	\\
2001	&	 4.58 	&	2	&	4	&	9	&	16	&	46	&	369	\\
2002	&	 4.23 	&	1	&	3	&	8	&	15	&	44	&	359	\\
2003	&	 4.31 	&	2	&	4	&	9	&	15	&	45	&	508	\\
2004	&	 4.43 	&	2	&	4	&	9	&	16	&	45	&	586	\\
2005	&	 4.45 	&	2	&	4	&	9	&	16	&	45	&	609	\\
2006	&	 4.44 	&	2	&	4	&	9	&	16	&	45	&	597	\\
2007	&	 4.49 	&	2	&	4	&	9	&	16	&	47	&	592	\\
2008	&	 4.39 	&	2	&	4	&	9	&	16	&	45	&	595	\\ \hline \hline
    \end{tabular}
    \begin{tabular}{c|c|c|c|c|c}
    \multicolumn{6}{c}{ } \\
    \multicolumn{6}{c}{Share of Exported Value by Foreign Suppliers by Distribution of \# Products} \\ \hline \hline
	&	Bottom 40\%	&	40\% to 60\%	&	60\% to 80\%	&	Top 20\%	&	Top 5\%	\\ \hline \hline
2000	&	 13.28 	&	 7.62 	&	 13.17 	&	 65.93 	&	46.27	\\
2001	&	 13.57 	&	 8.62 	&	 13.42 	&	 64.39 	&	45.82	\\
2002	&	 17.39 	&	 8.47 	&	 9.79 	&	 64.35 	&	43.94	\\
2003	&	 15.60 	&	 8.23 	&	 13.96 	&	 62.21 	&	45.35	\\
2004	&	 14.26 	&	 8.28 	&	 12.84 	&	 64.62 	&	47.26	\\
2005	&	 14.11 	&	 8.95 	&	 13.10 	&	 63.84 	&	48.15	\\
2006	&	 13.81 	&	 7.47 	&	 12.02 	&	 66.71 	&	51.34	\\
2007	&	 13.36 	&	 7.91 	&	 12.04 	&	 66.69 	&	49.41	\\
2008	&	 14.53 	&	 8.67 	&	 11.88 	&	 64.93 	&	49.16	\\
 \hline \hline
    \end{tabular}
\end{table}
\begin{table}[ht]
    \centering
    \caption{Foreign Suppliers' Distribution of Export Source Countries}
    \label{tab:fs_dist_source}
    \tiny
    \begin{tabular}{c|c|c|c|c|c|c|c}
    \multicolumn{8}{c}{ } \\
    \multicolumn{8}{c}{Number of Export Source Countries per Foreign Supplier} \\ \hline \hline
	&	Mean	&	Median	&	p75	&	p90	&	p95	&	p99	&	Max	\\
2000	&	1.34	&	1	&	1	&	2	&	3	&	7	&	36	\\
2001	&	1.34	&	1	&	1	&	2	&	3	&	7	&	38	\\
2002	&	1.39	&	1	&	1	&	2	&	3	&	8	&	35	\\
2003	&	1.38	&	1	&	1	&	2	&	3	&	8	&	40	\\
2004	&	1.38	&	1	&	1	&	2	&	3	&	8	&	42	\\
2005	&	1.38	&	1	&	1	&	2	&	3	&	8	&	43	\\
2006	&	1.38	&	1	&	1	&	2	&	3	&	8	&	47	\\
2007	&	1.39	&	1	&	1	&	2	&	3	&	8	&	49	\\
2008	&	1.40	&	1	&	1	&	2	&	3	&	8	&	51	\\ \hline \hline
    \end{tabular}
    \begin{tabular}{c|c|c|c}
    \multicolumn{4}{c}{ } \\
    \multicolumn{4}{c}{Share of Exported Value by Foreign Suppliers by Distribution of \# Source Countries} \\ \hline \hline
	&	Bottom 80\%	&	Top 20\%	&	Top 5\%	\\ \hline \hline
2000	&	 39.03 	&	 60.97 	&	 45.03 	\\
2001	&	 41.45 	&	 58.55 	&	 43.87 	\\
2002	&	 39.77 	&	 60.23 	&	 42.36 	\\
2003	&	 40.32 	&	 59.68 	&	 44.25 	\\
2004	&	 39.11 	&	 60.89 	&	 47.41 	\\
2005	&	 37.38 	&	 62.62 	&	 48.15 	\\
2006	&	 36.77 	&	 63.23 	&	 49.63 	\\
2007	&	 34.93 	&	 65.07 	&	 49.80 	\\ 
2008	&	 34.44 	&	 65.56 	&	 50.23 	\\ \hline \hline
    \end{tabular}    
\end{table}
\begin{table}[ht]
    \centering
    \caption{Foreign Suppliers' Distribution of Customers}
    \label{tab:fs_dist_customers}
    \tiny
    \begin{tabular}{c|c|c|c|c|c|c|c}
    \multicolumn{8}{c}{ } \\
    \multicolumn{8}{c}{Number of Export Source Countries per Foreign Supplier} \\ \hline \hline
	&	Mean	&	Median	&	p75	&	p90	&	p95	&	p99	&	Max	\\ \hline \hline
2000	&	 1.49 	&	1	&	1	&	2	&	4	&	9	&	125	\\
2001	&	 1.45 	&	1	&	1	&	2	&	3	&	8	&	107	\\
2002	&	 1.36 	&	1	&	1	&	2	&	3	&	7	&	65	\\
2003	&	 1.43 	&	1	&	1	&	2	&	3	&	9	&	86	\\
2004	&	 1.43 	&	1	&	1	&	2	&	3	&	8	&	85	\\
2005	&	 1.43 	&	1	&	1	&	2	&	3	&	9	&	94	\\
2006	&	 1.42 	&	1	&	1	&	2	&	3	&	8	&	102	\\
2007	&	 1.41 	&	1	&	1	&	2	&	3	&	8	&	116	\\
2008	&	 1.40 	&	1	&	1	&	2	&	3	&	8	&	108	\\ \hline \hline
    \end{tabular}
    \begin{tabular}{c|c|c|c}
    \multicolumn{4}{c}{ } \\
    \multicolumn{4}{c}{Share of Exported Value by Foreign Suppliers by Distribution of \# Customers} \\ \hline \hline
	&	Bottom 80\%	&	Top 20\%	&	Top 5\%	\\ \hline \hline
2000	&	45.60	&	54.40	&	34.36	\\
2001	&	48.36	&	51.64	&	36.50	\\
2002	&	48.33	&	51.67	&	34.33	\\
2003	&	47.33	&	52.67	&	37.46	\\
2004	&	46.87	&	53.13	&	37.37	\\
2005	&	43.31	&	56.69	&	40.40	\\
2006	&	43.54	&	56.46	&	41.80	\\
2007	&	45.11	&	54.89	&	40.85	\\
2008	&	43.37	&	56.63	&	42.54	\\  \hline \hline
    \end{tabular}    
\end{table}

Tables \ref{tab:fs_dist_instances}-\ref{tab:fs_dist_source} present this information. Each table has two panels. The top panel presents statistics on the distribution of the number of ``domestic Argentinean firm''-``product HS-6''-``source country'' level, at the ``product HS-6'' level, at the ``source country'' level, and at the ``domestic Argentinean firm'' level (or customer level) per foreign supplier. The bottom panel constructs the share of total exported value by foreign suppliers, i.e., total value imported by Argentinean firms, explained by firms categorized across different percentiles of the distribution of firms in the top panel. For instance, in the bottom panel of Table \ref{tab:fs_dist_instances}, the first column shows how much of the total value traded is explained by firms in the ``Bottom 40\%'' of the distribution of ``Number of Export Instances per Foreign Supplier''. The second column shows how much of the total value traded is explained by firms in the ``40\% to 60\%'' percentiles of the distribution of ``Number of Export Instances per Foreign Supplier'', etc.

Across Tables \ref{tab:fs_dist_instances}-\ref{tab:fs_dist_customers} it is clear that foreign suppliers are significantly heterogeneous across each other. In particular, the top-panels of these tables suggest that the distribution of the number of ``domestic Argentinean firm''-``product HS-6''-``source country'' level, at the ``product HS-6'' level, at the ``source country'' level, and at the ``domestic Argentinean firm'' level (or customer level) per foreign supplier are highly skewed towards higher values. Furthermore, the bottom panels of these tables suggest that foreign suppliers which export more products, from more source countries, and to more Argentinean customers explain a vast share of Argentina's total import flows.

\newpage
\noindent
\textbf{Granularity of Foreign Suppliers.} Next, I present evidence of the granular nature of aggregate imported value. To do so, I carry out a statistical exercise inspired by the work of \cite{gabaix2011granular}. I construct a measure of the top $K$ foreign supplier specific idiosyncratic shock, which is called \textit{granular residual}:
\begin{align*}
    \Gamma_t = \sum^K_{i=1} \frac{S_{i,t-1}}{Y_{t-1}} \left(g_{i,t} - \bar{g}^Q_t\right)
\end{align*}
where $S_{i,t-1}/Y_{t-1}$ is the share of foreign supplier $i$ in Argentina's aggregate import flows in year $t-1$, and $\left(g_{i,t} - \bar{g}^Q_t\right)$ is the demeaned growth rate of imported value, considered as a proxy for the idiosyncratic shock to supplier $i$ in year $t$. I follow \cite{blanco2018determination} and construct $\bar{g}^Q_t$ as the cross-sectional median growth rate of imported value across suppliers, i.e. $g_{i,t}$, computed among the top $Q$ firms, with $Q \geq K$. 

Following \cite{gabaix2011granular}, I estimate the regression
\begin{align*}
    g^Y_t = \alpha + \beta_i \Gamma_t + \epsilon_t
\end{align*}
where $g^Y_t$ is the aggregate growth rate of value imported, and use the $R^2$ to assess to which extent idiosyncratic foreign supplier shocks account for total value imported fluctuations. Table \ref{tab:granularity_aggregate} presents the results of this exercise for different values of $K$.
\begin{table}[ht]
    \centering
    \caption{Granularity of Foreign Suppliers in Aggregate Imported Value}
    \label{tab:granularity_aggregate}
    \begin{tabular}{c c c c}
                   & $Q=K =$ 50 & $Q=K =$ 100 & $Q=K =$ 200\\ \hline \hline
      $R^2$        & 0.6284 & 0.8942 & 0.9634 \\ 
      $\bar{R}^2$  & 0.5665 &  0.8766 &  0.9573 \\ \hline \hline
    \end{tabular}
\end{table}
Overall, this table suggests that a vast majority of aggregate imported value fluctuations are explained idiosyncratic shocks to the top foreign suppliers. Idiosyncratic shocks to the top 50 foreign suppliers explain between 56\% and 63\% of total aggregate import fluctuations. Idiosyncratic shocks to the top 200 foreign suppliers explain above 96\% of total aggregate import fluctuations.

While this result is constructed using a short temporal sample, it provides additional evidence of the granular nature of imports in a small open economy. This is, both at the aggregate and firm level, idiosyncratic foreign supplier shocks have a significant impact. 

\noindent
\textbf{Coverage of Employment.} In Sections \ref{sec:data_facts} and \ref{sec:supplier_shocks} I use data on firms' level of employment to characterize firms' network of foreign suppliers and to carry out sample partition exercises. The source of this data is SIPA or ``Sistema Integrado Previsional Argentino'' and the time lapse is 2001-2008. The coverage of the dataset on firm level employment differs across time. I turn to presenting descriptive statistics about firms' employment and to measure how representative these data set is.

First, Table \ref{tab:coverage_firm_employment} presents statistics on the number of firms in the firm level employment data set, the number of total employed workers in the data set, and the total number of registered private sector formal workers downloaded from the Argentina's Ministry of Labor website \small \url{https://www.trabajo.gob.ar/estadisticas/index.asp} \normalsize. 
\begin{table}[ht]
    \centering
    \footnotesize
    \caption{Coverage of Firms and Employment}
    \label{tab:coverage_firm_employment}
    \begin{tabular}{c|c|c|c}
	&	Number of Firms	&	Number of Employees	&	Argentina Private Formal Workers	\\ \hline \hline
    2001	&	211,456	&	 4,249,313 	&	 3,919,869 	\\
    2002	&	201,612	&	 3,922,558 	&	 3,522,917 	\\
    2003	&	209,221	&	 4,142,832 	&	 3,728,064 	\\
    2004	&	220,729	&	 4,536,320 	&	 4,182,778 	\\
    2005	&	231,251	&	 4,975,690 	&	 4,692,031 	\\
    2006	&	241,360	&	 5,405,175 	&	 5,137,830 	\\
    2007	&	614,145	&	 6,908,385 	&	 5,584,609 	\\
    2008	&	637,351	&	 7,291,401 	&	 5,923,320 \\ \hline \hline
    \end{tabular}
\end{table}
The discrepancies between the total number of Employees in the middle and right columns of Table \ref{tab:coverage_firm_employment} are driven by the presence of some public sector firms in the firm level employment data set. However, these firms are part of the service sector and do not participate in export activities.

In order to test whether the coverage of my employment level data set is sufficient for the study of import and export performance, I compute how much of total imports and exports are explained by firms in the employment dataset.
\begin{table}[ht]
    \centering
    \caption{Coverage of Employment Data Set \\ \footnotesize For Total Exports}
    \label{tab:employment_export}
    \footnotesize
    \begin{tabular}{c|c|c|c|c}
	& Not in Data Set	&	In Data Set	&	Not in Data Set	&	In Data Set	\\ \hline \hline
	&	\multicolumn{2}{c}{Value Exported} & \multicolumn{2}{c}{Share of Total in \%} \\ \hline \hline
2001	&	 724,178,289 	&	 26,313,288,048 	&	2.68	&	97.32	\\
2002	&	 431,464,547 	&	 25,164,527,098 	&	1.69	&	98.31	\\
2003	&	 445,095,525 	&	 28,935,173,075 	&	1.51	&	98.49	\\
2004	&	 568,011,356 	&	 33,754,234,558 	&	1.65	&	98.35	\\
2005	&	 738,143,360 	&	 38,834,260,689 	&	1.87	&	98.13	\\
2006	&	 798,389,056 	&	 45,465,646,096 	&	1.73	&	98.27	\\
2007	&	 564,533,438 	&	 54,797,513,124 	&	1.02	&	98.98	\\
2008	&	 756,060,037 	&	 68,723,011,281 	&	1.09	&	98.91	\\ \hline \hline
    \end{tabular}
\end{table}
\begin{table}[ht]
    \centering
    \caption{Coverage of Employment Data Set \\ \footnotesize For Total Imports}
    \label{tab:employment_import}
    \footnotesize
    \begin{tabular}{c|c|c|c|c}
	& Not in Data Set	&	In Data Set	&	Not in Data Set	&	In Data Set	\\ \hline \hline
	&	\multicolumn{2}{c}{Value Exported} & \multicolumn{2}{c}{Share of Total in \%} \\ \hline \hline
2001	&	 596,136,733 	&	 18,485,436,561 	&	3.12	&	96.88	\\
2002	&	 195,370,908 	&	 8,098,319,418 	&	2.36	&	97.64	\\
2003	&	 318,386,676 	&	 12,649,890,937 	&	2.46	&	97.54	\\
2004	&	 430,335,287 	&	 20,467,280,889 	&	2.06	&	97.94	\\
2005	&	 647,816,819 	&	 26,346,058,013 	&	2.40	&	97.60	\\
2006	&	 548,217,222 	&	 31,778,548,552 	&	1.70	&	98.30	\\
2007	&	 592,412,061 	&	 41,848,288,618 	&	1.40	&	98.60	\\
2008	&	 796,608,632 	&	 53,778,020,449 	&	1.46	&	98.54	\\ \hline \hline
    \end{tabular}
\end{table}
Tables \ref{tab:employment_export} and \ref{tab:employment_import} show the share of total exports and total imports explained by firms in and out of the employment dataset. A vast majority, above 98\% on average, of total exported and imported value is explained by firms in the employment data set used in the paper. This assures that the analysis carried out in Sections \ref{sec:data_facts} and \ref{sec:supplier_shocks} is representative of Argentinean firms engaging in international trade. 

Finally, to complement the analysis, Table \ref{tab:employment_descriptive_statistics} present descriptive statistics on firm level employment for the period 2001-2008. 
\begin{table}[ht]
    \centering
    \caption{Descriptive Statistics of Firm Level Employment}
    \label{tab:employment_descriptive_statistics}
    \footnotesize
    \begin{tabular}{c|c|c|c|c|c|c|c}
	&	Mean	&	p25	&	p50	&	p75	&	p90	&	p95	&	p99	\\ \hline \hline
2001	&	20.10	&	1	&	3	&	9	&	27	&	53	&	212.50	\\
2002	&	19.46	&	1	&	3	&	9	&	25	&	51	&	200.00	\\
2003	&	19.80	&	1	&	3	&	9	&	26	&	53	&	206.50	\\
2004	&	20.55	&	1	&	3	&	10	&	28	&	56	&	217.50	\\
2005	&	21.52	&	1	&	3	&	10	&	30	&	60	&	231.50	\\
2006	&	22.39	&	1	&	4	&	11	&	31	&	62	&	243.50	\\
2007	&	11.25	&	1	&	2	&	5	&	14	&	28	&	122.25	\\
2008	&	11.44	&	1	&	2	&	5	&	14	&	28	&	122.58	\\ \hline \hline
    \end{tabular}
\end{table}

Finally, I present evidence on the survival profile of domestic firms in the domestic market. 
\begin{figure}[ht]
    \centering
    \includegraphics[scale=0.5]{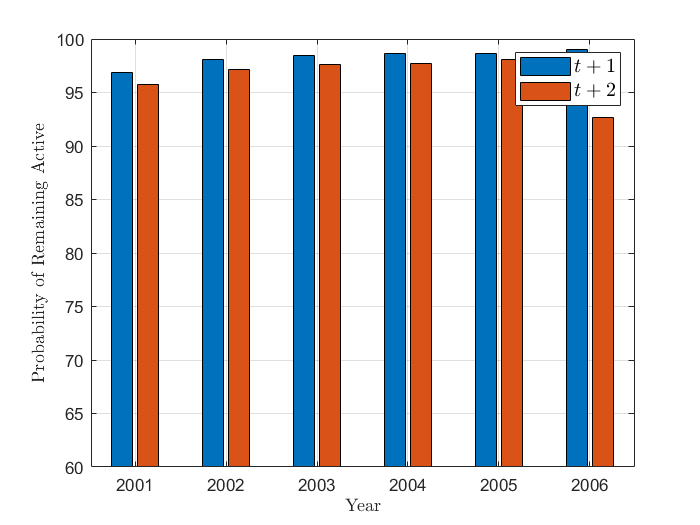}
    \caption{Survival Profile in the Domestic Market}
    \label{fig:survival_profile_domestic}
\end{figure}
Figure \ref{fig:survival_profile_domestic} shows the probability of a firm active in period $t$ of remaining active in periods $t+1$ and $t+2$. Across the time period, the average probability of remaining active in period $t+1$ and $t+2$ is 96\% and 92\% respectively. These are significantly higher survival probabilities than those experienced by the linkages between domestic and foreign firms studied in Section \ref{subsec:search_evidence}.

%%%%%%%%%%%%%%%%%%%%%%%%%%%%%%%%%%%%%%%%%%%%%%%%%%%%%%%%%%%%%
\newpage
\section{Supplier-Supply Shocks: Descriptive Statistics} \label{sec:appendix_shocks_desc}

In this section of the Appendix I present descriptive statistics about the Supplier-Supply-Shocks constructed in Section \ref{subsec:supplier_supply_shocks}. Table \ref{tab:appendix_shock_mean} presents the mean and standard deviation of the constructed shocks across the sample.
\begin{table}[ht]
    \centering
    \caption{Supplier-Supply Shocks Descriptive Statistics \\ \footnotesize Mean \& St. Dev.}
    \footnotesize
    \label{tab:appendix_shock_mean}
    \begin{tabular}{c c c}
    Year & Mean & Standard Deviation  \\ \hline \hline
    2001 & -0.489\% & 17.574\% \\
    2002 & -0.125\% & 12.808\% \\
    2003 & -0.374\% & 15.275\% \\
    2004 & -0.212\% & 15.520\% \\
    2005 & 0.013\% & 17.690\% \\
    2006 & 0.261\% & 19.674\% \\
    2007 & 0.108\% & 19.273\% \\
    2008 & -0.126\% & 19.918\% \\ \hline
    \textbf{Total Sample} & -0.059\% & 16.882\% \\ \hline \hline
    \end{tabular}
\end{table}

\begin{figure}[ht]
    \centering
    \caption{Auto-Correlation of Supplier-Supply Shocks}
    \label{fig:histogram_corr_shocks}
    \includegraphics[scale=0.45]{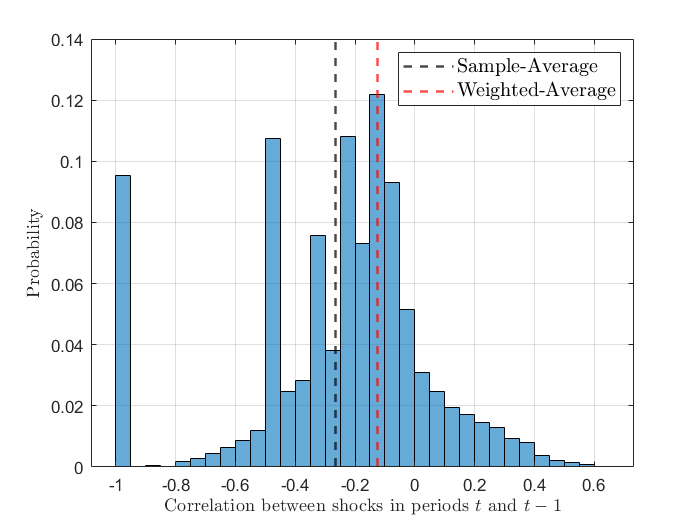}
\end{figure}

\noindent
Figure \ref{fig:histogram_corr_shocks} describes the auto-correlation of the constructed shocks. 
The average and weighted average of the correlations are slightly negative, implying that these shocks have an auto-regressive nature. Implying that a Supplier-Supply-Shock has a transitory nature.

Table \ref{tab:appendix_shock_mean_second_specification} presents the mean and standard deviation of the second specification of the Supplier-Supply-Shock constructed in Appendix \ref{appendix_regression_results}. 
\begin{table}[ht]
    \centering
    \caption{Supplier-Supply Shocks - Second Shock Specification \\ \footnotesize Mean \& St. Dev.}
    \label{tab:appendix_shock_mean_second_specification}
    \footnotesize
    \begin{tabular}{c c c}
    Year & Mean & Standard Deviation  \\ \hline \hline
        2001	&	-0.134\%	&	8.405\%	\\
    2002	&	-0.107\%	&	6.768\%	\\
    2003	&	0.127\%	&	7.547\%	\\
    2004	&	0.329\%	&	8.727\%	\\
    2005	&	0.450\%	&	8.534\%	\\
    2006	&	0.386\%	&	9.219\%	\\
    2007	&	0.512\%	&	9.565\%	\\
    2008	&	0.734\%	&	10.266\% \\ \hline
    \textbf{Total Sample} & 0.22557\% & 8.4716\% \\ \hline \hline
    \end{tabular}
\end{table}
The average shock exhibits a full sample mean of 0.226\% and a standard deviation of 8.4716\%. 

%%%%%%%%%%%%%%%%%%%%%%%%%%%%%%%%%%%%%%%%%%%%%%%%%%
\newpage
\section{Role of Importing in Exporting} \label{appendix_role_importing}

In this appendix I present additional evidence on the importance of imports on export performance. First, the regression analysis presented in Section \ref{sec:supplier_shocks} only focuses on exporting firms which imports. Thus, to assert that the results are important for aggregate exports as a whole it is necessary to show the relevance of importing in exporting.

\begin{figure}[ht]
    \caption{Importance of Imports for Exporting}
    \label{fig:import_for_exporting}
     \centering
     \begin{subfigure}[b]{0.495\textwidth}
         \centering
         \includegraphics[width=\textwidth]{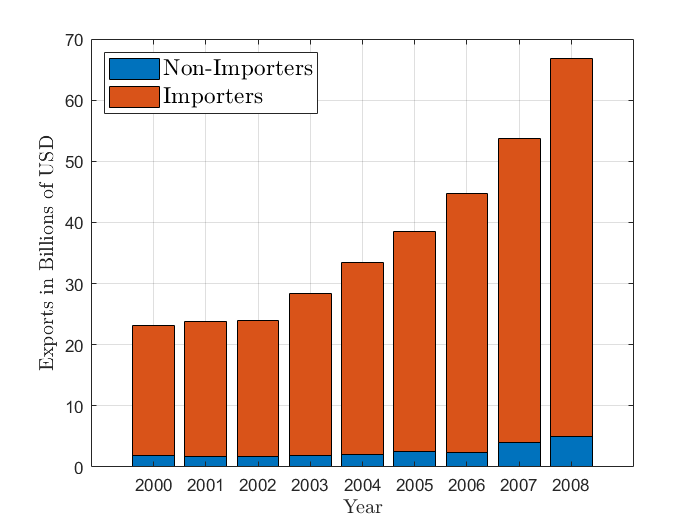}
         \caption{Share of Exports by Importers}
         \label{fig:Export_Decomposition_Importers}
     \end{subfigure}
     \hfill
     \begin{subfigure}[b]{0.495\textwidth}
         \centering
         \includegraphics[width=\textwidth]{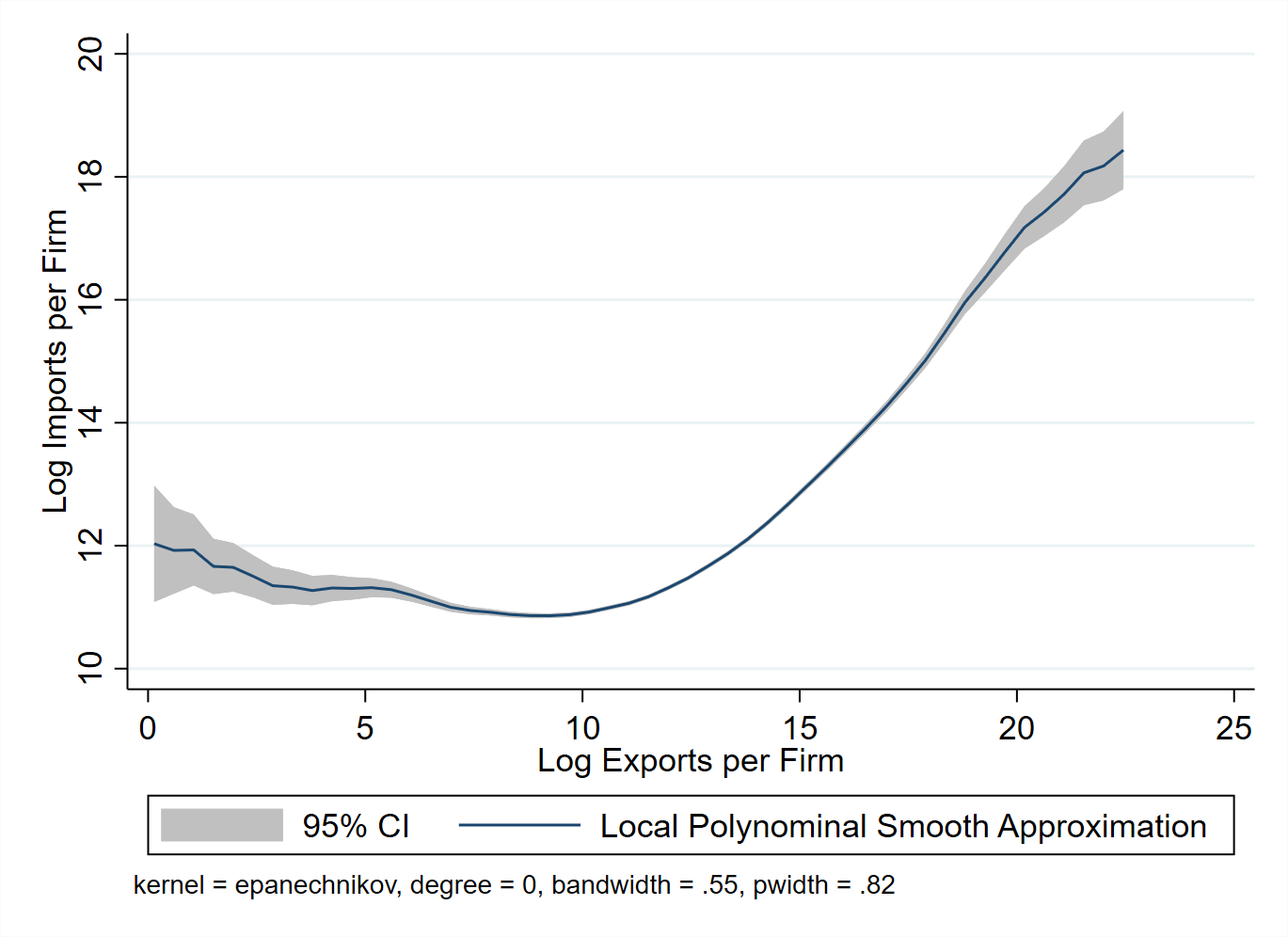}
         \caption{Exports \& Imports per Firm}
         \label{fig:exporter_importer_prime}
     \end{subfigure}
%    \floatfoot{\footnotesize \textbf{Note:} }
\end{figure}

Figure \ref{fig:import_for_exporting} presents two pieces of information which highlight the relevance of imports for exporting.
The left panel, Figure \ref{fig:Export_Decomposition_Importers} shows that a vast majority of total exported value is explained by importer firms (in red), with non exporter firms (in blue) explaining less than 7\% of annual exports. The right panel, Figure \ref{fig:exporter_importer_prime} shows the relationship between imports and exports at the firm level, through a kernel-weighted local polynomial smoothing fit of the data. This figure shows a strong relationship between exports and imports at the firm level, especially for firms exporting more than USD 10,000.

%%%%%%%%%%%%%%%%%%%%%%%%%%%%%%%%%%%%%%%%%%%%%%%%%%
\newpage
\section{Additional Regression results} \label{appendix_regression_results}

In this appendix of the paper I present additional evidence that support the regression results presented in Sections \ref{subsec:import_performance} and \ref{subsec:export_performance}. 

First, Table \ref{tab:export_instances} presented the impact of a supplier-supply shock on exported quantities. Column (1) showed that estimating Equation \ref{eq:regression_exports} without controlling for lagged values of exported quantities or the supplier-supply shock leads to results not significantly different from zero. However, when controlling for these lagged values results are significantly different from zero.
\begin{figure}[ht]
    \centering
    \caption{Importance of Imports in Exports in Regression Results}
    \label{fig:importance_regression_exports}    
    \includegraphics[scale=0.5]{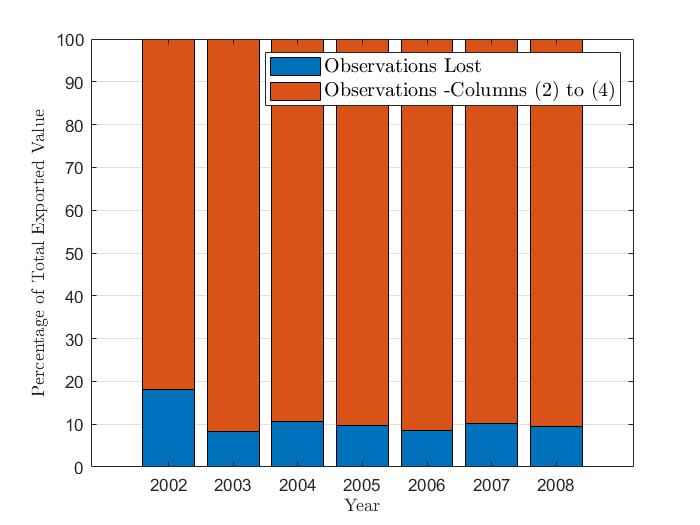}
\end{figure}
Figure \ref{fig:importance_regression_exports} shows that while controlling for lagged values of these variables reduces the number of observations of the regression exercise, these lost observations represent on average 10\% of the total value exported implied by the observations in column (1). Thus, columns (2) through (4) show that a supplier-supply shock leads to a significant negative decrease in exported quantities for a vast majority of total exported value.

Next, I present robustness regression results. In particular, instead of using the log change in prices in the construction of Supplier-Supply-Shocks, I use the percentage difference
\begin{align} \label{eq:change_prices_percentage_differences_appendix}
    \Delta  \tilde{P}_{i,k,t} = \frac{ P_{i,k,t} - P_{i,k,t-1}}{0.5 \times \left(P_{i,k,t} + P_{i,k,t-1} \right)}
\end{align}
I re-estimate equation \ref{eq:price_reg_supplier} as
\begin{align} \label{eq:price_reg_supplier_percentage_differences_appendix}
    \Delta \tilde{P}_{i,p,d,s,t} = \tilde{\beta}_{i,t} + \tilde{\gamma}_{s,t} + \tilde{\epsilon}_{i,p,d,s,t}
\end{align}
and construct the new Supplier-Supply-Shocks
\begin{align}
    \tilde{\text{Supplier-Supply-Shock}}_{i,t} = \sum_s \omega_{s,t-1} \times \tilde{\gamma}_{s,t}
\end{align}

\noindent
\textbf{Robustness checks import performance results.} I start by showing the robustness of the impact a Supplier-Supply Shock on imported quantities at the $\{i,p,d,s\}$ level under the new shock specification. Table \ref{tab:import_instances_supplier_pdiff} presents these results.
\begin{table}[ht]
    \centering
    \caption{Supplier-Supply Shock \& Imported Quantities \\ \small Robustness Check - $\Delta \tilde{P}_{i,k,t}$}
    \label{tab:import_instances_supplier_pdiff}
    \footnotesize
\begin{tabular}{lcccc} 
 & \multicolumn{4}{c}{Imported Quantities - $\ln Q_{i,p,d,s,t}$} \\
 & (1) & (2) & (3) & (4) \\ \hline \hline
\\
$\text{Supplier-Supply-Shock}_{i,t}$ & -0.150*** & -0.259*** & -0.252*** & -0.253*** \\
 & (0.0219) & (0.0348) & (0.0331) & (0.0311) \\
 &  &  &  &  \\
$\ln Q_{i,p,d}$ lags & NO & YES & YES & YES \\
$\text{Supplier-Supply-Shock}_{i}$ lags & NO & YES & YES & YES \\
$\ln$ RER$_{d,t}$ \& lags & NO & NO & YES & YES \\
$\ln$ Total imported value & NO & NO & NO & YES \\  
 &  &  &  &  \\
Observations & 2,098,170 & 650,785 & 541,475 & 541,475 \\ \hline \hline
% R-squared & 0.914 & 0.925 & 0.927 & 0.927 \\ \hline
\multicolumn{5}{c}{ Two-way clustered standard errors in parentheses (Firm-Source country level)} \\
\multicolumn{5}{c}{ *** p$<$0.01, ** p$<$0.05, * p$<$0.1} \\
\end{tabular}
\floatfoot{\footnotesize \textbf{Note:} Every specification in the table controls for $\gamma_{i,p,d}$ and $\gamma_t$ fixed effects. The specifications which control for lagged values of either the explained and/or the explanatory variables do so by controlling for the $t-1$ and $t-2$ values of the variables. }
\end{table}
As expected, results are robust and statistically different from zero. Table \ref{tab:import_instances_supplier_pdiff_empleo_total} shows that the results by level of employment are also robust.
\begin{table}[ht]
    \centering
    \caption{Supplier-Supply Shock on Imported Quantities \\ \footnotesize By Employment Levels - $\Delta \tilde{P}_{i,k,t}$}
    \label{tab:import_instances_supplier_pdiff_empleo_total}
    \footnotesize
\begin{tabular}{l c c c} 
 & \multicolumn{3}{c}{Imported Quantities - $\ln QX_{i,p,d,s,t}$} \\
 & \textbf{Total} & \textit{Above} & \textit{Below} \\
 & (1) & (2) & (3) \\ \hline \hline
\\
$\tilde{\text{Supplier-Supply-Shock}}_{i,t}$ & -0.353*** & -0.316* & -0.394*** \\
 & (0.0815) & (0.172) & (0.116) \\
 &  &  &  \\
$\ln Q_{i,p,d}$ lags & YES & YES & YES \\
$\text{Supplier-Supply-Shock}_{i}$ lags & YES & YES & YES \\
$\ln$ RER$_{d,t}$ \& lags & YES & YES & YES \\
$\ln$ Total imported value & YES & YES & YES \\  
 &  &  &  \\
Observations & 159,803 & 100,387 & 53,132 \\\hline \hline
% R-squared & 0.914 & 0.925 & 0.927 & 0.927 \\ \hline
\multicolumn{4}{c}{ Two-way clustered standard errors in parentheses (Firm-Destination country level)} \\
\multicolumn{4}{c}{ *** p$<$0.01, ** p$<$0.05, * p$<$0.1} \\
\end{tabular}
\floatfoot{\footnotesize \textbf{Note:} Every specification in the table controls for $\gamma_{i,p,d}$ and $\gamma_t$ fixed effects. The specifications which control for lagged values of either the explained and/or the explanatory variables do so by controlling for the $t-1$ and $t-2$ values of the variables. In order to partition firms between \textit{Above} and \textit{Below} I compute the mean level of employment \textit{Above}-\textit{Below} threshold at the ``\textit{year} - \textit{production sector} (at the 6 digit level)'' for firms actively importing. }
\end{table}
Relatively smaller firms have a larger estimated impact and more statistically different from zero than relatively larger firms. These results are in line with the results presented in Section \ref{subsec:import_performance}. 

Next, I turn to estimating the impact of the newly constructed import cost shock on quantities imported at the $\{i,p,d\}$ level. Table \ref{tab:import_instances_pdiff} shows the impact of the newly constructed Supplier-Supply Shock on imported quantities at the $\{i,p,d\}$.
\begin{table}[ht]
    \centering
    \caption{Supplier-Supply Shock on Imported Quantities \\ \footnotesize Robustness Check - $\Delta \tilde{P}_{i,k,t}$}
    \label{tab:import_instances_pdiff}
    \footnotesize
\begin{tabular}{lcccc} 
 & \multicolumn{4}{c}{Imported Quantities - $\ln Q_{i,p,d,t}$} \\
 & (1) & (2) & (3) & (4) \\ \hline \hline
\\
$\text{Supplier-Supply-Shock}_{i,t}$ & -0.106*** & -0.159*** & -0.153*** & -0.153*** \\
 & (0.0237) & (0.0311) & (0.0259) & (0.0261) \\
 &  &  &  &  \\
$\ln Q_{i,p,d}$ lags & NO & YES & YES & YES \\
$\text{Supplier-Supply-Shock}_{i}$ lags & NO & YES & YES & YES \\
$\ln$ RER$_{d,t}$ \& lags & NO & NO & YES & YES \\
$\ln$ Total imported value & NO & NO & NO & YES \\  
 &  &  &  &  \\
Observations & 1,971,366 & 836,056 & 699,143 & 699,143 \\ \hline \hline
% R-squared & 0.914 & 0.925 & 0.927 & 0.927 \\ \hline
\multicolumn{5}{c}{ Two-way clustered standard errors in parentheses (Firm-Source country level)} \\
\multicolumn{5}{c}{ *** p$<$0.01, ** p$<$0.05, * p$<$0.1} \\
\end{tabular}
\floatfoot{\footnotesize \textbf{Note:} Every specification in the table controls for $\gamma_{i,p,d}$ and $\gamma_t$ fixed effects. The specifications which control for lagged values of either the explained and/or the explanatory variables do so by controlling for the $t-1$ and $t-2$ values of the variables. }
\end{table}
As expected, results are robust, with the estimated coefficient being negative and statistically different from zero.

Lastly, I test whether the results are robust across levels of employment. Table \ref{tab:import_instances_pdiff_empleo_total} show that as in the benchmark results, relatively smaller firms show estimated impacts twice as big as the benchmark, while relatively larger firms are unaffected. 
\begin{table}[ht]
    \centering
    \caption{Supplier-Supply Shock \& Imported Quantities \\ \footnotesize By Employment Levels - $\Delta \tilde{P}_{i,k,t}$}
    \label{tab:import_instances_pdiff_empleo_total}
    \footnotesize
\begin{tabular}{l c c c} 
 & \multicolumn{3}{c}{Imported Quantities - $\ln Q_{i,p,d,t}$} \\
 & \textbf{Total} & \textit{Above} & \textit{Below} \\
 & (1) & (2) & (3) \\ \hline \hline
\\
$\tilde{\text{Supplier-Supply-Shock}}_{i,t}$ & -0.224*** & -0.179 & -0.358*** \\
 & (0.0514) & (0.120) & (0.0811) \\
 &  &  &  \\
$\ln Q_{i,p,d}$ lags & YES & YES & YES \\
$\text{Supplier-Supply-Shock}_{i}$ lags & YES & YES & YES \\
$\ln$ RER$_{d,t}$ \& lags & YES & YES & YES \\
$\ln$ Total imported value & YES & YES & YES \\  
 &  &  &  \\
Observations & 211,213 & 130,936 & 70,806 \\ \hline \hline
% R-squared & 0.914 & 0.925 & 0.927 & 0.927 \\ \hline
\multicolumn{4}{c}{ Two-way clustered standard errors in parentheses (Firm-Destination country level)} \\
\multicolumn{4}{c}{ *** p$<$0.01, ** p$<$0.05, * p$<$0.1} \\
\end{tabular}
\floatfoot{\footnotesize \textbf{Note:} Every specification in the table controls for $\gamma_{i,p,d}$ and $\gamma_t$ fixed effects. The specifications which control for lagged values of either the explained and/or the explanatory variables do so by controlling for the $t-1$ and $t-2$ values of the variables. In order to partition firms between \textit{Above} and \textit{Below} I compute the mean level of employment \textit{Above}-\textit{Below} threshold at the ``\textit{year} - \textit{production sector} (at the 6 digit level)'' for firms actively importing. }
\end{table}

\noindent
\textbf{Robustness checks export performance results.} Next, I turn to using the newly constructed shocks $\tilde{\text{Supplier-Supply-Shock}}_{i,t}$  and test the robustness to the results presented in Section \ref{subsec:export_performance}. Table \ref{tab:export_instances_pdiff_appendix} shows that the results presented in Table \ref{tab:export_instances} are robust to the new shock specification.
\begin{table}[ht]
    \centering
    \caption{Supplier-Supply Shock \& Exported Quantities \\ \footnotesize Robustness Check - $\Delta \tilde{P}_{i,k,t}$}
    \label{tab:export_instances_pdiff_appendix}
    \small
\begin{tabular}{lcccc} 
 & \multicolumn{4}{c}{Exported Quantities - $\ln QX_{i,p,d,t}$} \\
 & (1) & (2) & (3) & (4) \\ \hline \hline
\\
$\tilde{\text{Supplier-Supply-Shock}}_{i,t}$ & -0.00913 & -0.145*** & -0.134*** & -0.136*** \\
 & (0.0279) & (0.0497) & (0.0502) & (0.0500) \\
 &  &  &  &  \\
$\ln Q_{i,p,d}$ lags & NO & YES & YES & YES \\
$\tilde{\text{Supplier-Supply-Shock}}_{i}$ lags & NO & YES & YES & YES \\
$\ln$ RER$_{d,t}$ \& lags & NO & NO & YES & YES \\
$\ln$ Total imported value & NO & NO & NO & YES \\  
 &  &  &  &  \\
Observations & 415,371 & 153,534 & 137,696 & 137,696 \\ \hline \hline
% R-squared & 0.914 & 0.925 & 0.927 & 0.927 \\ \hline
\multicolumn{5}{c}{ Two-way clustered standard errors in parentheses (Firm-Destination country level)} \\
\multicolumn{5}{c}{ *** p$<$0.01, ** p$<$0.05, * p$<$0.1} \\
\end{tabular}
\floatfoot{\footnotesize \textbf{Note:} Every specification in the table controls for $\gamma_{i,p,d}$ and $\gamma_t$ fixed effects. The specifications which control for lagged values of either the explained and/or the explanatory variables do so by controlling for the $t-1$ and $t-2$ values of the variables. }
\end{table}
As in the case for the benchmark shock specification, the results are not robust when not controlling for lagged values of the shock and of the dependent variable, but highly significant for all other specifications.

Now, I turn to showing that the results presented in Table \ref{tab:export_instances_empleo_total} are robust to the new shock specification.
\begin{table}[ht]
    \centering
    \caption{Supplier-Supply Shock \& Exported Quantities \\ \footnotesize By Employment Level - $\Delta \tilde{P}_{i,k,t}$}
    \label{tab:export_instances_pdiff_empleo_total_appendix}
    \small
\begin{tabular}{l c c c} 
 & \multicolumn{3}{c}{Exported Quantities - $\ln QX_{i,p,d,t}$} \\
 & \textbf{Total} & \textit{Above} & \textit{Below} \\
 & (1) & (2) & (3) \\ \hline \hline
\\
$\tilde{\text{Supplier-Supply-Shock}}_{i,t}$ & -0.149*** & -0.0789 & -0.215*** \\
 & (0.0482) & (0.0644) & (0.0783) \\
 &  &  &  \\
$\ln Q_{i,p,d}$ lags & YES & YES & YES \\
$\text{Supplier-Supply-Shock}_{i}$ lags & YES & YES & YES \\
$\ln$ RER$_{d,t}$ \& lags & YES & YES & YES \\
$\ln$ Total imported value & YES & YES & YES \\  
 &  &  &  \\
Observations & 118,397 & 75,421 & 38,175 \\ \hline \hline
% R-squared & 0.914 & 0.925 & 0.927 & 0.927 \\ \hline
\multicolumn{4}{c}{ Two-way clustered standard errors in parentheses (Firm-Destination country level)} \\
\multicolumn{4}{c}{ *** p$<$0.01, ** p$<$0.05, * p$<$0.1} \\
\end{tabular}
\floatfoot{\footnotesize \textbf{Note:} Every specification in the table controls for $\gamma_{i,p,d}$ and $\gamma_t$ fixed effects. The specifications which control for lagged values of either the explained and/or the explanatory variables do so by controlling for the $t-1$ and $t-2$ values of the variables. In order to partition firms between \textit{Above} and \textit{Below} I compute the mean level of employment \textit{Above}-\textit{Below} threshold at the ``\textit{year} - \textit{production sector} (at the 6 digit level)'' for firms actively exporting. }
\end{table}
Table \ref{tab:export_instances_pdiff_empleo_total_appendix} presents the results of this robustness check. In line with the benchmark results, the estimated impact on relatively smaller firms is significantly larger than the full sample estimate. Furthermore, as shown for the benchmark results, the impact on relatively larger firms is not significantly different from zero.

%%%%%%%%%%%%%%%%%%%%%%%%%%%%%%%%%%%%%%%%%%
\newpage
\section{Proofs} \label{sec:appendix_proofs}

In this section of the paper I present the proofs to Proposition 1 in Section \ref{sec:model}.

\noindent
\textit{Proof of Proposition 1:} Firm $i$'s production line $j$ profit maximization problem can be stated as
\begin{equation*}
\begin{aligned}
\max_{x_{i,j},p_{i,j}} \pi_{i,j} \left(z_i,\tau_j\right) = \quad & \left(p_{i,j} - \frac{\tau_j}{z_{i}} C_{i,j} \right) x_{i,j} - e_jF^{e} \mathbbm{1}\left[x_{i,j}>0\right] \\ %- f |\Omega_{i,j}|^{\mu} \\
\textrm{s.t.} \quad & x_{i,j} = p^{-\rho}_{i,j} Y_j P^{\rho}_j \\
  &  C_{i} \left(w, p^M_{i} \right) = \left(\alpha^{\frac{1}{1-\theta}} w^{\frac{\theta}{\theta-1}} + \left(1-\alpha\right)^{\frac{1}{1-\theta}} \left(p^M_{i}\right)^{\frac{\theta}{\theta-1}}\right)^{\frac{\theta-1}{\theta}} \\
  &  p^M_{i} = \left[\sum^{K}_{k=1} p^{\frac{\varphi}{\varphi-1}}_{i,k} \right]^{\frac{\varphi-1}{\varphi}}
\end{aligned}
\end{equation*}
Using the demand schedule on the first constraint to substitute $p_{i,j}$ the profit maximization problem can be written only in terms of $x_{i,j}$
\begin{equation*}
\begin{aligned}
\max_{x_{i,j}} \pi_{i,j} \left(z_i,\tau_j\right) = \quad & x^{1-\frac{1}{\rho}} P_j Y^{\frac{1}{\rho}}_{j} - \frac{\tau_j}{z_i} C_{i} x_{i,j} \\ 
\textrm{s.t.} \quad & x_{i,j} = p^{-\rho}_{i,j} Y_j P^{\rho}_j \\
  &  C_{i} \left(w, p^M_{i} \right) = \left(\alpha^{\frac{1}{1-\theta}} w^{\frac{\theta}{\theta-1}} + \left(1-\alpha\right)^{\frac{1}{1-\theta}} \left(p^M_{i}\right)^{\frac{\theta}{\theta-1}}\right)^{\frac{\theta-1}{\theta}} \\
  &  p^M_{i} = \left[\sum^{K}_{k=1} p^{\frac{\varphi}{\varphi-1}}_{i,k} \right]^{\frac{\varphi-1}{\varphi}}
\end{aligned}
\end{equation*}
The first order condition with respect to $x_{i,j}$ is given by
\begin{align*}
    \left(1-\frac{1}{\rho}\right)x^{\frac{-1}{\rho}}_{i} P_j Y^{\frac{1}{\rho}}_j = \frac{\tau_j}{z_i} C_i
\end{align*}
Manipulating the FOC above, I can obtain the optimal quantities produced 
\begin{align*}
    x^{*}_{i,j} = \left(\frac{\tau_j}{z_i} C_i \right)^{-\rho} P^{\rho}_j Y_j \left(\frac{\rho}{\rho-1}\right)^{-\rho}
\end{align*}
Replacing the subscript $j$ with $f$ leads to the optimal quantities exported presented in Equation \ref{eq:optimal_quantities_exported}. I can compute firm $i$'s profits by plugging the optimal quantities produced into the profit maximization objective function for production line $f$
\begin{align*}
    \pi^{*}_{i,j} \left(z_i,\tau_f\right) &=  \left(x^{*}\right)^{1-\frac{1}{\rho}} P_j Y^{\frac{1}{\rho}}_{j} - \frac{\tau_j}{z_i} C_{i} \left(x^{*}\right) \\
    &= \left(\frac{\tau_f}{z_i}C_i\right)^{1-\rho} P^{\rho} Y_f \left(\frac{\rho}{\rho-1}\right)^{-\rho} \left(\frac{1}{\rho-1}\right)
\end{align*}
In order to obtain threshold $\bar{z}$ such that firm $i$ exports if $z_i>\bar{z}$, I introduce the fixed cost of exporting $F^{e}$ in terms of labor and set firm $i$'s profits to zero
\begin{align*}
    \left(\frac{\tau_f}{\bar{z}}C_i\right)^{1-\rho} P^{\rho} Y_f \left(\frac{\rho}{\rho-1}\right)^{-\rho} \left(\frac{1}{\rho-1}\right) - w F^e = 0
\end{align*}
Manipulating this expression and solving for $\bar{z}$ yields
\begin{align*}
    \bar{z} = \frac{\tau_f C_{i}}{\rho-1} \left[ \rho^{\rho} \frac{w F^e}{P^{\rho}_f Y_f} \right]^{\frac{1}{\rho-1}} 
\end{align*}
which is exactly the expression in Equation \ref{eq:export_threshold}.

\newpage
\section{Details on Model's Features \& Calibration} \label{sec:appendix_model_calibration_fit}

In this Appendix of the paper I present additional details on the model's features, simulation and its calibration fit. 

\noindent
\textbf{Computation details.} The calibration and simulation exercises are carried out by simulating the path of a large number of firms. In particular, I simulate idiosyncratic firm level productivities $z_i$ by taking 5,000 draws from a log-normal distribution with mean $\mu_z$ and $\sigma_z$. These are time invariant productivities.

Note that Equation \ref{eq:fixed_cost_function} implies that searching for the first foreign supplier is cost-less. Thus, I simulate 5,000 draws from a uniform distribution between $\left[\underline{p},\bar{p}\right]$ and allocate it as firms' first foreign suppliers' efficiency.

Next, for firm $i$, given her idiosyncratic productivity $z_i$ and initial foreign supplier efficiency $p^{1}$, I simulate her search and matching process until she decides to stop searching. To do this, I compute condition
\begin{align*} \label{eq:search_condition_appendix}
    \frac{\beta}{1-\beta} \times \pi^S\left(z_i,\{p_{i,k}\}^K_{k=1}\right) \geq F^{S}\left(K\right)
\end{align*}
If the condition is met, I take a draw from $\left[\underline{p},\bar{p}\right]$ and expand firm $i$'s set of foreign suppliers. In the following period, I repeat the exercise until 
\begin{align*} 
    \frac{\beta}{1-\beta} \times \pi^S\left(z_i,\{p_{i,k}\}^K_{k=1}\right) < F^{S}\left(K\right)
\end{align*}
which implies that the firm does not search anymore. After the firm stops searching, I move to simulating firm $i+1$'s set of foreign suppliers.

Once the computation of all 5,000 firms' set of foreign suppliers is done I can compute firms' profits, exported quantities and imports of intermediate inputs. Furthermore, I can compute aggregate statistics.

\noindent
\textbf{Calibration details.} In order to assign numerical values for the vectors of parameters $\Theta = \{f^s,\mu,\bar{p}\}$ and $\{\sigma_z, F^{e}\}$ I seek to minimize the difference between a set of moments from the data and their model counterparts. In particular, I follow an approach similar to \cite{mcfadden1989method}'s simulated method of moments. For a given set of parameters $\{f^s,\mu,\bar{p},\sigma_z, F^{e}\}$ I carry out the simulation of firms' supplier sets, total imported value of intermediate inputs, and export decisions based on the algorithm described above. From these simulations I compute the model counterpart of the data moments described in Section \ref{subsec:model_numerical}. I compute the difference between data and model moments. I square the difference between each of these moments and sum them. I use a non-linear numerical solver to minimize this sum of square differences.

\noindent
\textbf{Figures of sensitivity analysis.} Next, I present the figures referred in Section \ref{subsec:model_numerical} which describe the sensitivity of the model's aggregate moments as a function of $\{f^{s},\mu,\bar{p},\varphi\}$. 
\begin{figure}
    \centering
    \includegraphics[scale=0.65]{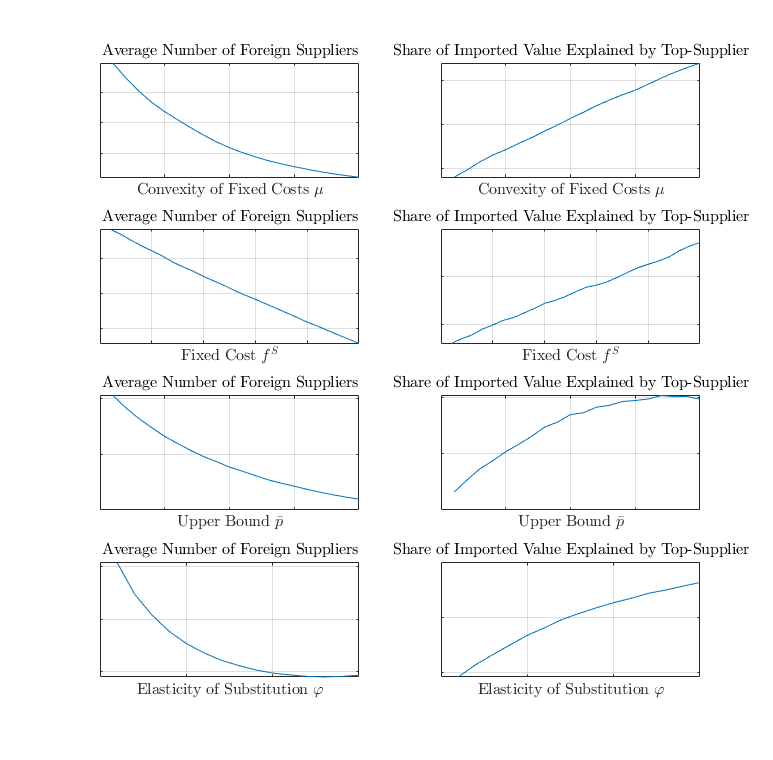}
    \caption{Aggregate Moments \& Model Parameters \\ \footnotesize Sensitivity Analysis}
    \label{fig:Sensitivity_Analysis}
\end{figure}
Figure \ref{fig:Sensitivity_Analysis} shows how the average number of suppliers and the mean share of imported value by the top-supplier change as a function of these parameters. 

\begin{figure}[ht]
    \centering
    \caption{Model's Supplier Supply Shock \\ \footnotesize Sensitivity Analysis }
    \label{fig:IRFs_sensitivity}
    \begin{subfigure}[b]{0.45\textwidth}
    \centering    
    \includegraphics[scale=0.4]{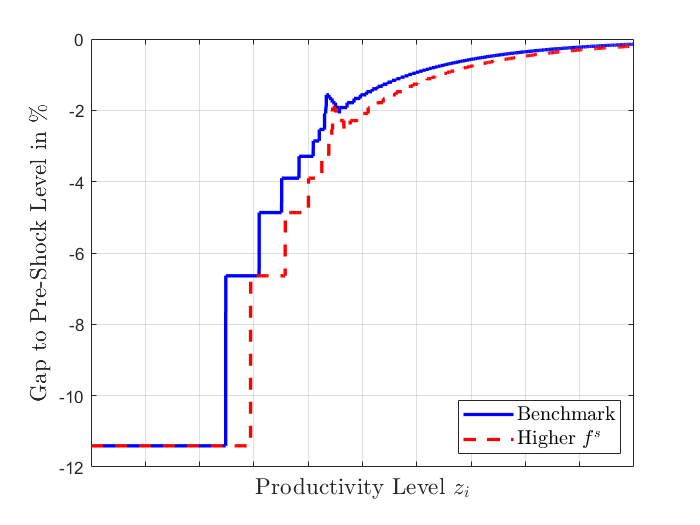}
    \caption{Higher $f^{s}$}
    \label{fig:IRFs_top_exports_fixed_cost}
     \end{subfigure}
     \hfill
    \begin{subfigure}[b]{0.45\textwidth}
    \centering
    \includegraphics[scale=0.4]{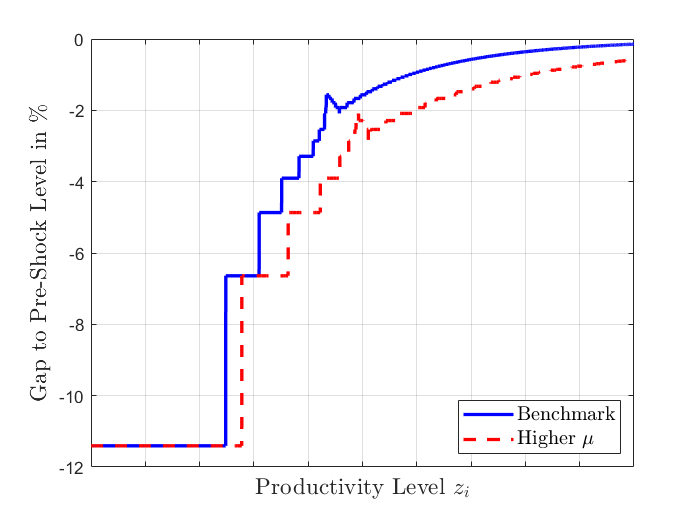}
    \caption{Higher $\mu$}
    \label{fig:IRFs_top_exports_high_convexity}
     \end{subfigure} \\
    \begin{subfigure}[b]{0.45\textwidth}
    \centering    
    \includegraphics[scale=0.4]{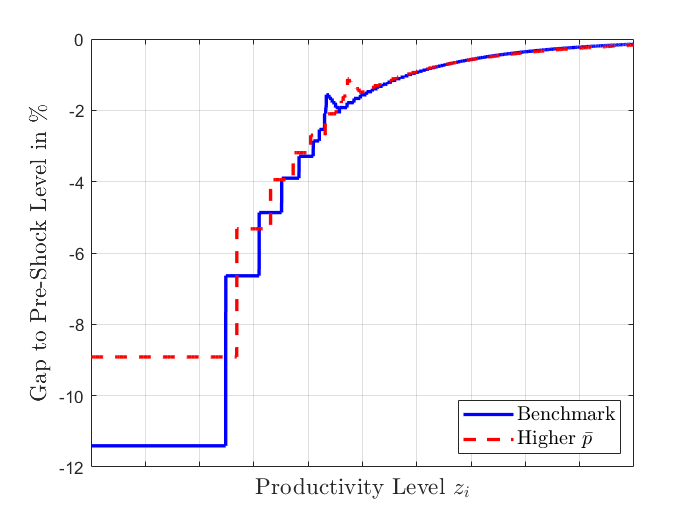}
    \caption{Higher $\bar{p}$}
    \label{fig:IRFs_top_exports_high_ub}
     \end{subfigure}
     \hfill
    \begin{subfigure}[b]{0.45\textwidth}
    \centering
    \includegraphics[scale=0.4]{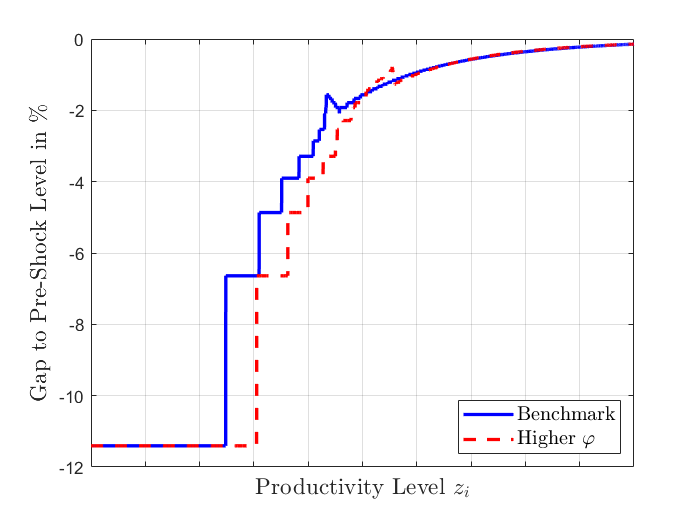}
    \caption{Higher $\varphi$}
    \label{fig:IRFs_top_exports_high_varphi}
     \end{subfigure}
\end{figure}

Figure \ref{fig:IRFs_sensitivity} presents how differences in parameters $\{f^{s},\mu,\bar{p},\varphi\}$ affect the impact of a Supplier-Supply shock on firms' export performance. 
The results are described in Section \ref{subsec:model_numerical}.

\begin{figure}[ht]
    \centering
    \includegraphics[scale=0.5]{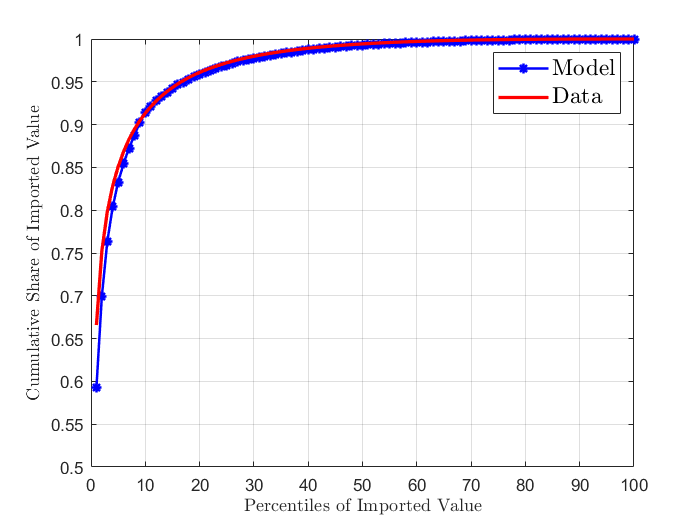}
    \caption{Model's Fit - Distribution of Imported Value per Firm}
    \label{fig:cum_share_impo}
\end{figure}

\noindent
\textbf{Additional measures of fit.} First, Figure \ref{fig:cum_share_impo} presents the fit of the model's implied distribution of imported value by firm with its data counterpart. 
This shows that the model calibration, particular of parameters $\sigma_z$, provides an adequate representation of the distribution of firms' imports.

\end{document}